\documentstyle[emulateapj,apjfonts,colordvi]{article}

\newcommand{\Msun}{$M_{\odot}$}
\newcommand{\kms}{km~s$^{-1}$} 
\def\h2{H$_{2}$}
\def\fh2{$f_{\rm H2}$}

\newcommand{\ebv}{E(B-V)}

\def\n01{$N_{01}$} 
\def\t01{$T_{01}$} 
 
\def\nd{\nodata}
\input epsf

\begin{document}

\title{A FUSE SURVEY OF INTERSTELLAR MOLECULAR HYDROGEN IN THE SMALL
AND LARGE MAGELLANIC CLOUDS\altaffilmark{1}}
\author{JASON~TUMLINSON \altaffilmark{2}, J.~MICHAEL~SHULL\altaffilmark{2,3},
BRIAN~L.~RACHFORD\altaffilmark{2}, 
MATTHEW~K.~BROWNING\altaffilmark{2}, 
THEODORE~P.~SNOW\altaffilmark{2}, 
ALEX~W.~FULLERTON\altaffilmark{4,5}, 
EDWARD~B.~JENKINS\altaffilmark{6}, 
BLAIR~D.~SAVAGE\altaffilmark{7}, 
PAUL~A.~CROWTHER\altaffilmark{8}, 
H.~WARREN~MOOS\altaffilmark{5}, 
KENNETH~R.~SEMBACH\altaffilmark{5}, 
GEORGE~SONNEBORN\altaffilmark{9}, 
\& DONALD~G.~YORK\altaffilmark{10}  
}

\altaffiltext{1}{This work is based on data
obtained for the Guaranteed Time Team by the NASA-CNES-CSA FUSE mission
operated by the Johns Hopkins University.  Financial support to U.S.
participants has been provided by NASA contract NAS5-32985.} 
\altaffiltext{2}{Center for Astrophysics and Space Astronomy, 
Department of Astrophysical and Planetary
Sciences, University of Colorado, Boulder, CO 80309}
\altaffiltext{3}{Also at JILA, University of Colorado and National
Institute of Standards and Technology} 
\altaffiltext{4}{Department of Physics and Astronomy, University of Victoria, Victoria,  
                     BC, V8W 3P6, Canada} 
\altaffiltext{5}{Department of Physics and Astronomy, Johns Hopkins University, 
                 Baltimore, MD 21218} 
\altaffiltext{6}{Department of Astrophysical Sciences, 
                 Princeton University, Princeton, NJ 08544}
\altaffiltext{7}{Department of Astronomy, University of Wisconsin, Madison, WI 53706} 
\altaffiltext{8}{Department of Physics and Astronomy, 
                University College London, Gower Street, London WC1E 6BT, UK}
\altaffiltext{9}{NASA Goddard Space Flight Center, Code 681, Greenbelt, MD 20771} 
\altaffiltext{10}{Department of Astronomy and Astrophysics, University of Chicago, 
                 Chicago, IL 60637} 
\begin{abstract}
We describe a moderate-resolution FUSE survey of \h2\ along 70 sight
lines to the Small and Large Magellanic Clouds, using hot stars as
background sources. FUSE spectra of 67\% of observed Magellanic Cloud
sources (52\% of LMC and 92\% of SMC) exhibit absorption lines from the
\h2\ Lyman and Werner bands between 912 and 1120 \AA.  Our survey is
sensitive to N(\h2) $\geq 10^{14}$ cm$^{-2}$; the highest column
densities are log N(\h2) = 19.9 in the LMC and 20.6 in the SMC.  We
find reduced \h2\ abundances in the Magellanic Clouds relative to the
Milky Way, with average molecular fractions $\langle f_{\rm H2} \rangle
= 0.010^{+0.005}_{-0.002}$ for the SMC and $\langle f_{\rm H2} \rangle
= 0.012^{+0.006}_{-0.003}$ for the LMC, compared with $\langle f_{\rm
H2} \rangle = 0.095$ for the Galactic disk over a similar range of
reddening. The dominant uncertainty in this measurement results
from the systematic differences between 21 cm radio emission and
Ly$\alpha$ in pencil-beam sight lines as measures of N(H~I). These results
imply that the diffuse \h2\ masses of the LMC and SMC are $8 \times
10^{6}$ \Msun\ and $2 \times 10^{6}$ \Msun, respectively, 2\% and 0.5\%
of the H~I masses derived from 21 cm emission measurements.  The LMC
and SMC abundance patterns can be reproduced in ensembles of model
clouds with a reduced \h2\ formation rate coefficient, $R \sim 3 \times
10^{-18}$ cm$^{3}$ s$^{-1}$, and incident radiation fields ranging from
10 - 100 times the Galactic mean value.  We find that these
high-radiation, low-formation-rate models can also explain the enhanced
N(4)/N(2) and N(5)/N(3) rotational excitation ratios in the Clouds.  We
use \h2\ column densities in low rotational states ($J$ = 0 and 1) to
derive kinetic and/or rotational temperatures of diffuse interstellar
gas, and find that the distribution of rotational temperatures is
similar to Galactic gas, with $\langle T_{01} \rangle = 82 \pm 21$ K
for clouds with N(\h2) $\geq$ 10$^{16.5}$ cm$^{-2}$.  There is only a
weak correlation between detected \h2\ and far-infrared fluxes as
determined by IRAS, perhaps due to differences in the survey
techniques.  We find that the surface density of \h2\ probed by our
pencil-beam sight lines is far lower than that predicted from the
surface brightness of dust in IRAS maps.  We discuss the implications
of this work for theories of star formation in low-metallicity
environments.  \end{abstract}

\section{INTRODUCTION}

Molecular hydrogen (\h2) must play a central role in our understanding
of interstellar chemistry, but little is known about the distribution
of diffuse \h2\ in the interstellar medium (ISM) of the Galaxy.  Major
uncertainties remain about its formation, destruction, and recycling
into dense clouds and stars (see the review by Shull \& Beckwith 1982).
Since the first detection of interstellar \h2\ (Carruthers 1970) numerous
studies ({\em Copernicus}, Savage et al.~1977, hereafter S77; ORFEUS,
Richter~2000; FUSE, Shull et al.~2000) using ultraviolet absorption
measurements of \h2\ have created a general view of \h2\ formation,
destruction, and excitation in the diffuse ISM.  However, we lack
information about how the abundance of \h2\ and its physical parameters
depend on the environmental conditions of interstellar gas, such as the
metallicity, dust content, and UV radiation field. The \h2\ molecule has
not been studied comprehensively in interstellar environments outside
the Galactic disk. Varying metallicity and UV radiation may modify the
molecular abundance, thereby affecting the interstellar chemistry that
triggers the star formation process.  In particular, the formation rate
of \h2\ in the diffuse ISM may depend on the composition and physical
state of the gas and grains in unknown fashion.

Most of our knowledge of \h2\ in the diffuse Galactic ISM comes
from ultraviolet absorption measurements of the Lyman and Werner
rotational-vibrational bands in the far ultraviolet (912 -- 1120 \AA).
The {\em Copernicus} satellite mapped out the distribution and properties
of \h2\ along sight lines to stars confined to the local Galactic disk
(Spitzer, Cochran, \& Hirshfeld 1974; S77), and thus to a narrow range
of gas properties. Despite this limitation, these studies uncovered
important correlations between \h2\ abundance and the environmental
conditions (dust and gas).  The basic volume limitation on studies of
\h2\ persisted until the advent of new instruments. The more sensitive
but short-lived instruments, HUT (Gunderson, Clayton, \& Green 1998) and
ORFEUS (Dixon, Hurwitz, \& Bowyer~1998; Ryu et al.~2000; Richter~2000),
have extended the range of \h2\ studies to selected distant sources in
the Galaxy and the Magellanic Clouds, and the IMAPS spectrograph (Jenkins
\& Peimbert 1997; Jenkins et al. 2000) obtained high-resolution \h2\
spectra of bright targets in diverse environments.

Molecular hydrogen is thought to form on interstellar dust grains when
atoms are adsorbed on the grain surface, chemically bond there, and then
are ejected from the grain surface (Hollenbach, Werner, \& Salpeter
1971).  Theoretical calculations, accounting for the probabilities of
atom adsorption, and molecule formation and ejection, place the volume
formation rate on grain surfaces at $n_{\rm H} n_{\rm HI} R$, where $R
\approx 3 \times 10^{-17}$ cm$^{3}$ s$^{-1}$ in typical interstellar
conditions, and where it is assumed that the total hydrogen density,
$n_{H}$, is proportional to the number density of grains (Hollenbach \&
McKee 1979).  Observations by {\em Copernicus} support this idea, with
an inferred formation rate coefficient $R = (1-3) \times 10^{-17}$
cm$^{3}$ s$^{-1}$ (Jura 1974).  Correlations of \h2\ and dust can
provide a key test of this theory, but the narrow range of dust and
gas properties in the Galactic disk accessible to {\em Copernicus}
prohibited an exploration of the dependence of molecule formation on
dust properties in a variety of environments.  The lower-metallicity
environments of the SMC and LMC (Welty et al.~1997; Welty et al.~1999)
allow us to probe \h2\ formation and destruction in physical and chemical
environments different from the Galaxy.  In particular, the 4- to 17-
times lower dust-to-gas ratios in the Magellanic Clouds (Koornneef 1982;
Fitzpatrick 1985) imply a smaller grain surface area per hydrogen atom
and a correspondingly lower efficiency of \h2\ formation.

The Magellanic Clouds are therefore a valuable nearby laboratory for
studying \h2\ in more distant, low-metallicity star forming regions and
QSO absorption-line systems with high H~I column density. Molecular
hydrogen has been detected in three damped Ly$\alpha$ systems with
N(\h2)/N(H~I) in the range 10$^{-6}$ to 10$^{-4}$ (Ge \& Bechtold 1999;
Ge, Bechtold, \& Kulkarni 2001; Petitjean, Srianand, \& Ledoux~2000;
Black, Chaffee, \& Foltz 1987), and limits have been placed on molecular
abundance in several other systems. Some of these systems are known
to have low metallicity and/or high UV radiation intensity relative to
the Galaxy. Detailed studies of \h2\ in the nearby, spatially resolved
Magellanic Clouds can serve as an important benchmark for comparison
with these high-redshift systems.

Here, we describe the first survey results on \h2\ in the ISM of the
Magellanic Clouds obtained with the {\em Far Ultraviolet Spectroscopic
Explorer} (FUSE) satellite.  Initial FUSE results on \h2\ in the Milky Way
(Shull et al.~2000) suggest that diffuse \h2\ is ubiquitous in the Galaxy.
A large fraction of FUSE sight lines through the Galactic disk and halo
exhibit absorption from the Lyman and Werner rotational-vibrational
bands, showing that the {\em Copernicus} results on \h2\ in the local
regions of the disk extend, in principle, to more distant regions of
the Galaxy.  The sensitivity of the FUSE spectrograph allows the use
of distant hot stars and extragalactic objects as background sources,
opening up more distant regions of the Galaxy and beyond to study of the
\h2\ molecule.  Our observations survey the \h2\ abundance and properties
in the low-metallicity regimes of the LMC and SMC, allowing us to model
the formation and destruction rates of \h2, and the density and radiation
field in the interstellar gas.

In \S~2 we describe the observations and the data reduction and analysis
schemes. In \S~3 we survey the results on column densities, excitation,
and other observed properties of the detected \h2, discuss correlations
with other gas and dust properties along the sight lines, and compare
these results to numerical cloud models. Section 4 discusses these
results and summarizes the general conclusions of this survey.  In this
paper we assume a distance of 60 kpc to the SMC and 50 kpc to the LMC.

\section{OBSERVATIONS AND ANALYSIS} 

\subsection{FUSE Observations} 

The observations reported here comprise all the available LMC and SMC
targets from FUSE Guaranteed Time Observations during Cycle 1 (up to
October 2000).  The FUSE mission and its instrumental capabilities are
described by Moos et al.~(2000) and Sahnow et al.~(2000).  The target
stars were selected for campaigns to study O, B, and Wolf-Rayet stars
(Program ID P117) and to examine hot gas in the Milky Way (MW) and
Magellanic Clouds (P103).  Seven of the SMC stars was chosen specifically
for \h2\ studies (P115).  These observations present the first opportunity
to study \h2\ throughout the Magellanic Clouds in a comprehensive fashion.
We include in the sample three SMC stars and one LMC star previously
analyzed by Shull et al.~(2000), the LMC star Sk -67 05, previously
analyzed by Friedman et al.~(2000) and the SMC star Sk 108, previously
analyzed by Mallouris et al.~(2001).  An atlas of Magellanic Clouds stars
observed with FUSE has been compiled by Danforth et al.~(2001).  We list
the target stars, dataset names, and observational parameters in Tables
1 and 2.  A summary of results appears in Tables 3 and 4, where we list
detected column densities N(\h2), or upper limits, for the program stars.
Tables 5 and 6 list individual rotational level column densities N(J)
for the LMC and SMC, respectively, as derived from the curve-of-growth
and/or line-profile fits (\S~2.2).

\begin{figure*}[t] 
\centerline{\epsfxsize=1.0\hsize{\epsfbox{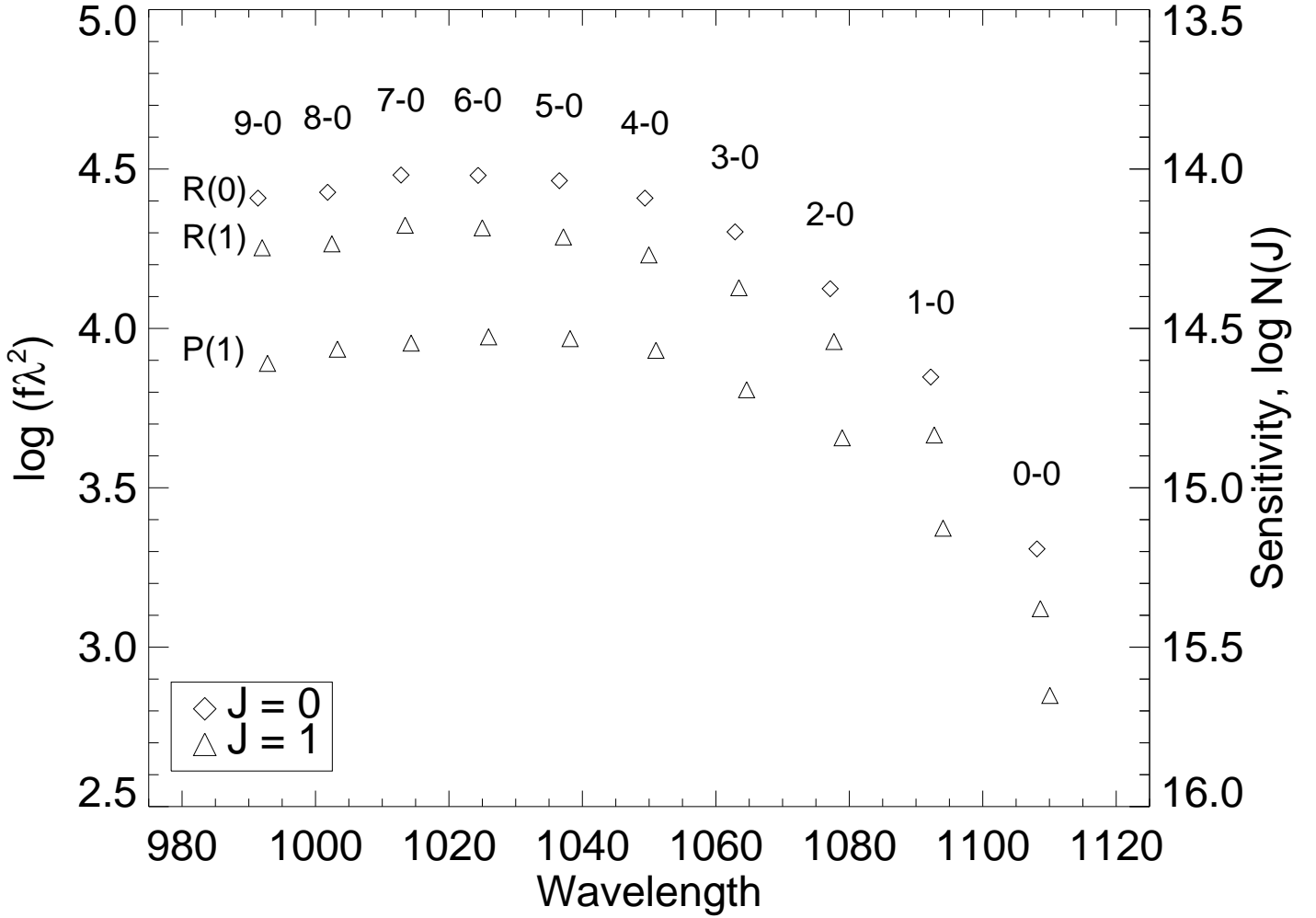}}}
\figcaption{\scriptsize The product $f\lambda ^{2}$ for the Lyman 0-0 to 9-0
bands of \h2. Oscillator strengths and wavelengths are from Abgrall et
al.~(1993a,b). The best constraints on the \h2\ curve of growth are provided
by the relatively weak 0-0 and 1-0 bands. The right axis shows 
the 4 $\sigma$ column density sensitivity for the lines displayed, assuming 
that the limiting equivalent width of all lines is 30 m\AA\ (see Equation 
1 and its discussion). 
\label{flfig}} 
\end{figure*}

All observations were obtained in time-tag (TTAG) mode through the
$30\arcsec \times 30\arcsec$ (LWRS) apertures.  Most observations were
broken into multiple exposures taken over consecutive orbital viewing
periods (see Tables 1 and 2).  The photon lists from these exposures
were concatenated before being processed by the current version of the
calibration pipeline software ({\sc calfuse} v1.8.7).  This program
computes the shifts in the detected position of a photon required to
correct for: (a) the motion of the satellite; (b) nodding motions of
the diffraction gratings, which are induced by thermal variations on an
orbital time scale; (c) small, thermally-induced drifts in the read-out
electronics of the detector; and (d) fixed geometric distortions in
the detector.  Application of these shifts produces a two-dimensional,
distortion-corrected image of a detector segment, from which a small,
uniform background is subtracted.  We extract one-dimensional spectra for
the LiF and SiC channels recorded on each of the four detector segments.
After correcting for detector dead time (which is small for count
rates typical of targets in the Magellanic Clouds), we apply the most
recent wavelength and effective-area calibrations to convert the detector
count rate in pixel space to flux units as a function of heliocentric
wavelength.  Throughout these processing steps, we propagate 1 $\sigma$
uncertainties along with the data.  No further modifications of the
{\sc calfuse}-processed data files are necessary for the {\h2}
analysis. 

The FUSE detectors have very low dark count rates, and the overall
level of light scattered to the detectors is low compared with the
fluxes of our targets. Our targets range in specific flux from
$F_{\lambda} = 5 \times 10^{-13}$ to $2 \times 10^{-12}$ erg cm$^{-2}$
s$^{-1}$ \AA$^{-1}$.  The typical background count rate is the sum of
dark counts and scattered light and corresponds to $< 10^{-14}$ erg
cm$^{-2}$ s$^{-1}$ \AA$^{-1}$. The {\sc calfuse} software performs a
background subtraction, making the contribution of the background to
the total error in continuum placement and equivalent widths
negligible.  

\begin{figure*}[t] 
\centerline{\epsfxsize=\hsize{\epsfbox{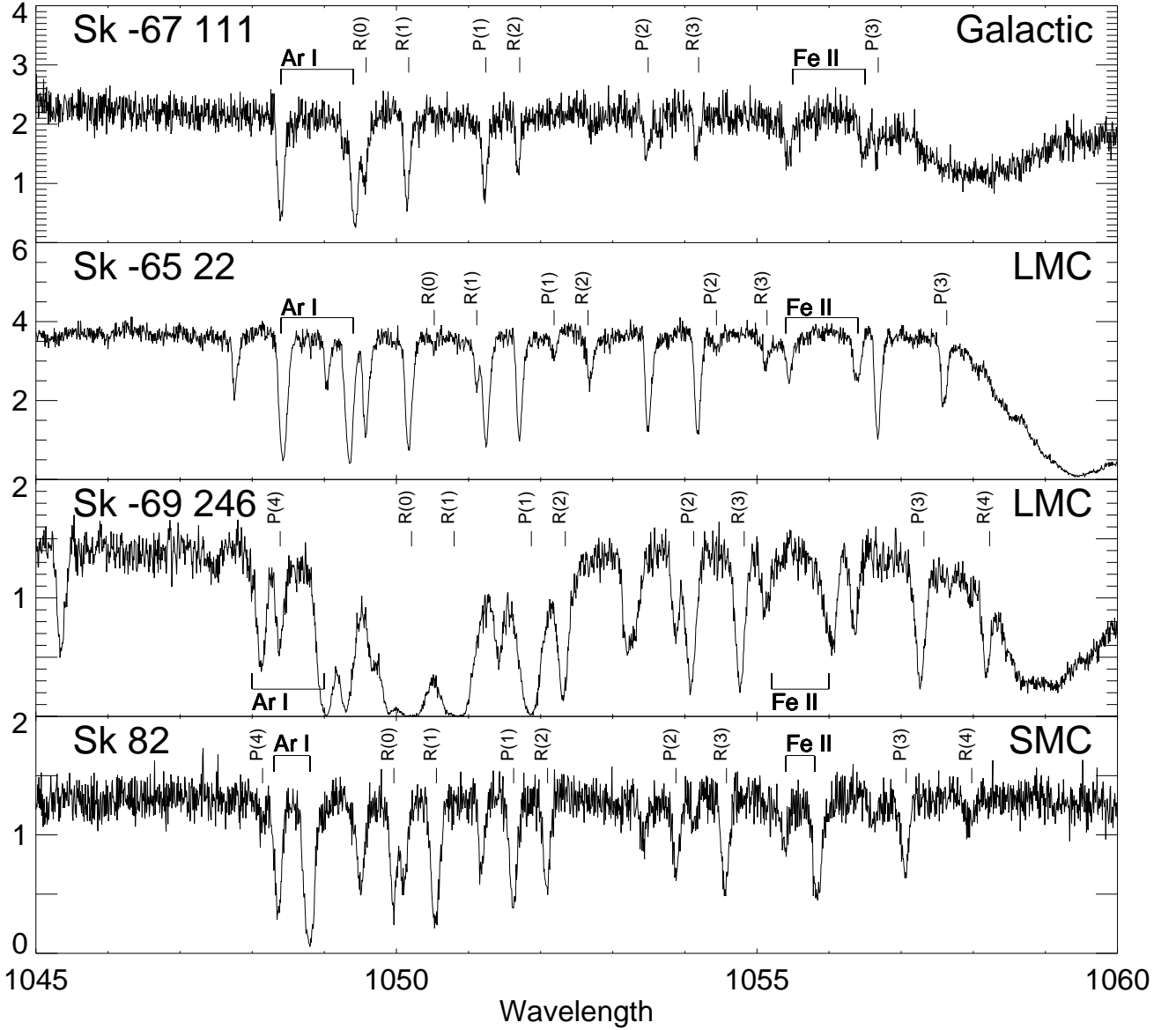}}}
\figcaption{\scriptsize Examples of FUSE spectra of SMC and LMC hot stars showing
varying column densities of \h2. First panel: Sk -67 111 shows no
detectable \h2 at LMC velocity, with N(\h2) $< 5.0 \times 10^{14}$
cm$^{-2}$. The Galactic \h2\ lines are marked here for reference in the
lower panels. Second panel: Sk -65 22, with N(\h2) = $7.6 \times 10^{14}$
cm$^{-2}$. Third panel: Sk -69 246, with N(\h2) = $6.3 \times 10^{19}$
cm$^{-2}$. Fourth panel: The SMC star Sk 82, with N(\h2) = $7.8 \times
10^{15}$ cm$^{-2}$.  The wavelength range is chosen to encompass the
Lyman 4-0 band of \h2. The Lyman 5-0 P(4) line appears at left in the
lower two panels. 
\label{specfig}}
\end{figure*}

Because FUSE contains no internal wavelength-calibration source, all
data are calibrated using a wavelength solution derived from in-orbit
observations of sources with well-studied interstellar components (Sahnow
et al.~2000). This process leads to relative wavelength errors of 10 -
20 km s$^{-1}$ across the band, in addition to a wavelength zero point
that varies between observations. Thus, we cannot rely completely on
radial velocity measurements of interstellar absorption.  In the analysis
below, we derive relative velocities from \h2\ lines and fix the zero
point in each sight line by assuming that the Milky Way absorption lies
at v$_{\rm LSR}$ = 0 km s$^{-1}$.


The resolution of the FUSE spectrograph across the band was $\lambda /
\Delta \lambda \simeq$ 10,000-20,000 for these observations.  Because
many of these observations were acquired during testing and calibration
phases of the FUSE mission, and because others were obtained late in
2000 after the instrument and its calibration had settled, the
instrumental resolution and other parameters vary from target to
target.  However, since the results rely on measured equivalent widths
and damping-profile fitting, our conclusions are not sensitive to the
resolution and spectrophotometric calibration.

\begin{figure*}[h] 
\centerline{\epsfxsize=\hsize{\epsfbox{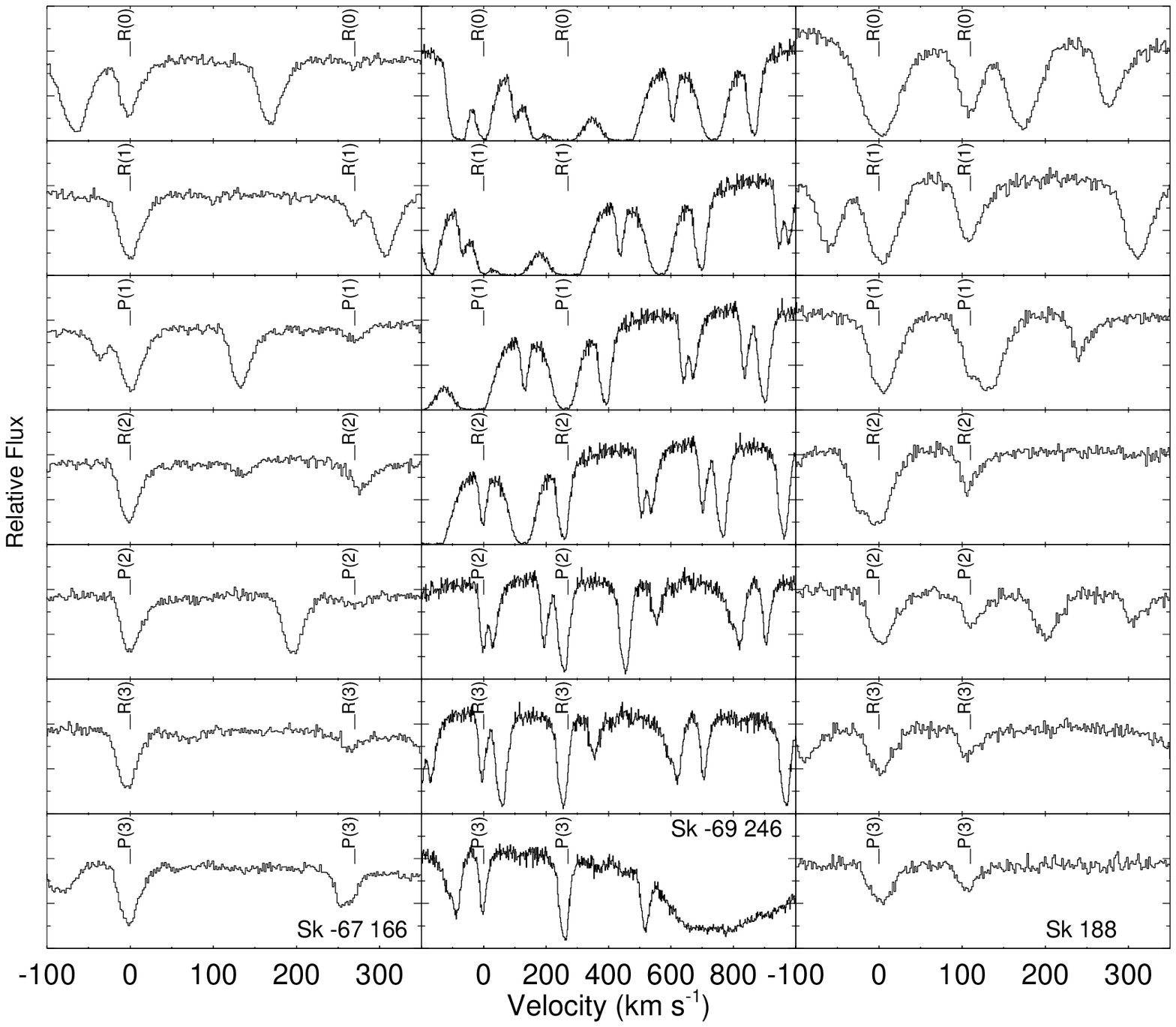}}}
\figcaption{\scriptsize Examples of the FUSE sample. The three stars shown are,
from left to right: Sk -67 166 (also seen in Figure 4) and Sk -69 246
in the LMC and Sk 188 in the SMC. The J=0, 1, and 2 lines of the Lyman
4-0 band are shown, plotted in velocity space. They are chosen to show
the velocity separations between Galactic and Magellanic \h2\ and to
bracket the range of detected \h2\ column densities. The left line in
each pair is the Milky Way component, the right line is the Magellanic
Cloud component.  The signal-to-noise ratios and resolution seen here
are typical of our sample. 
\label{velfig}}
\end{figure*}

\subsection{FUSE Data Analysis} 

In our analysis of \h2, we focus on the Lyman and Werner electronic
transitions, with some 400 vibrational-rotational lines arising from
the $J$ = 0~--~7 rotational states of the ground vibrational and
electronic states of \h2. We search for $J \leq 7$, but lines above
$J = 4$ are difficult to detect in diffuse interstellar clouds, given
the typical 4 $\sigma$ limiting equivalent width of 30 -- 40 m\AA\ and
the corresponding column density limit of $\rm N(H_2) \simeq 10^{14}$
cm$^{-2}$. In this analysis we use oscillator strengths, wavelengths,
and damping constants from Abgrall et al.~(1993a,b). The line strengths
$f\lambda^{2}$ for the R(0), R(1) and P(1) lines of the 0-0 to 9-0 Lyman
bands appear in Figure~\ref{flfig}. Starting at 0-0, the line strengths
increase by roughly a factor of ten to 7-0 and then decline slowly as
the upper vibrational level increases.

Figure~\ref{specfig} shows portions of FUSE spectra of four program
stars, with a range of column densities N(\h2).  We show the Lyman
(4-0) vibrational band and label Milky Way and Magellanic lines.
Absorption from gas in the Magellanic Clouds is easily distinguished by
a large velocity separation in the range v = 200 - 300 km
s$^{-1}$ for the LMC and 100 -- 170 km s$^{-1}$ for the SMC. This paper
is concerned with the Magellanic Cloud absorption only.  Results on the
detected Galactic gas in these sight lines will be presented later,
together with other FUSE targets at high Galactic latitude.

\begin{figure*}[t] 
\epsfxsize=\hsize{\epsfbox{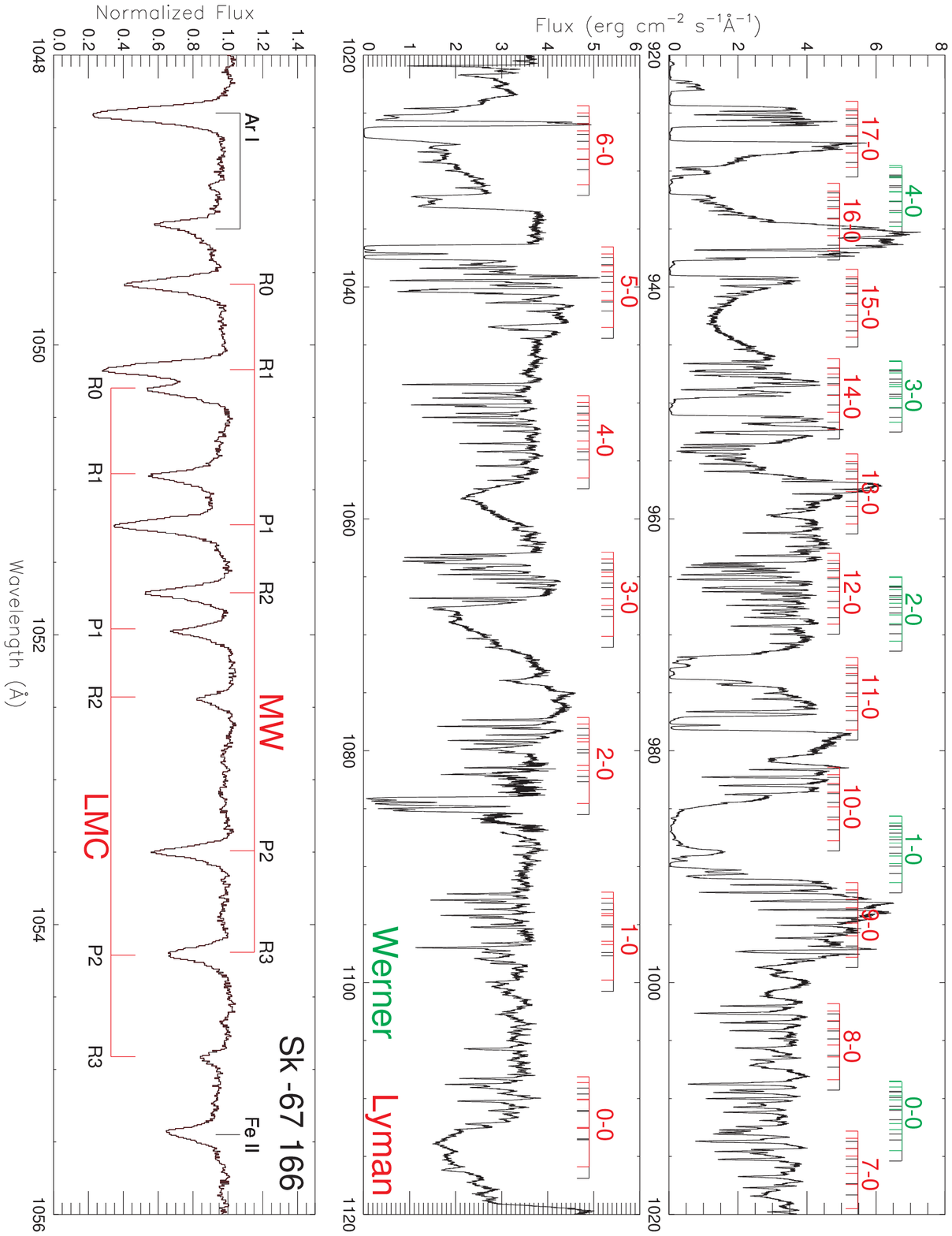}} 
\figcaption{\scriptsize Top two panels: FUSE spectrum of the LMC star Sk -67
166. This sight line has N(\h2) = $5.5 \times 10^{15}$ cm$^{-2}$ in J =
0 -- 3. The MW (black) and LMC (green and red) components are labeled
within each band. Lower panel: The Lyman 4-0 band of \h2. LMC and MW
lines are labeled. See the text for more discussion of this figure.
\label{fig}} 
\end{figure*}

Figure~\ref{velfig} shows the spectra of three selected stars, plotted in
velocity space to show the relative positions and strengths of the \h2\
lines of interest and the common blending between Galactic and Magellanic
components.  Figure~\ref{fig} shows the entire FUSE spectrum of one LMC
star, Sk -67 166. This figure illustrates the richness and complexity of
FUSE data on \h2. We have identified the Lyman (red) and Werner (green)
bands and the MW and LMC components. The LMC \h2\ component has N(\h2)
= $5.5 \times 10^{15}$ cm$^{-2}$ and the MW component has N(\h2) =
$5.0 \times 10^{15}$ cm$^{-2}$.

We employ a complement of techniques when analyzing \h2\ absorption
lines.  The final products of the analysis are measurements of the column
densities in the individual rotational levels, N(J), from which we can
infer the gas density, radiation field, and formation and destruction
rates of \h2.  For absorbers with $\rm N(H_{2}) \lesssim 10^{18}$
cm$^{-2}$, we measure equivalent widths, $W_{\lambda}$, of all \h2\ lines
and produce a curve of growth (COG) to infer a Doppler $b$-parameter and
N($J$).  For high-column density absorbers with damping wings in the J =
0 and 1 lines, we fit the line profiles to derive N(0) and N(1), and we
use a curve-of-growth technique to derive column densities for $J \geq
2$.  However, there is no well-defined column density limit above which
profile-fitting must be performed. Instead, it is used in any case for
which no lines of J~=~0 and 1 can be fitted with Gaussian line profiles, 
due to the presence of damping wings. 
 
We are often forced to neglect \h2\ lines in spectral regions near
strong interstellar absorption or bright geocoronal emission.  The \h2\
Lyman (6-0) band is always lost in the strong damping wings of the
interstellar Ly$\beta$ line (1025.7 \AA), and the Lyman (5-0) band lies
among the resonance absorption lines \ion{C}{2} $\lambda$1036.3 and
\ion{C}{2}$^{*}$ $\lambda$1037.0.  We neglect these bands except when
their $J \geq 2$ lines appear apart from the intervening absorption.

The key aspect of our analysis software is the rapid and consistent
measurement of as many individual \h2\ absorption lines as possible in
each sight line.  Variations in the data quality, line-of-sight structure,
and spectral type of the source prevent the use of a uniform set of \h2\
lines in the analysis.  Thus, the process of measuring the numerous
lines that enter the curve-of-growth fit cannot be automated completely.
Our software requires the user to decide on a line-by-line basis which
\h2\ lines will be fitted.

First, model spectra of varying N(J) and $b$ are overlaid on the
spectrum to obtain an estimate of the total column density N(\h2) and
the radial velocities of the Galactic and Magellanic components. At
this stage, detections are discriminated from non-detections and
analyzed separately.  We describe our analysis scheme for detections
here and return to discuss non-detections below.  Our software scans
for \h2\ lines from a list  and queries the user at each line
position.  If the user decides the line is present and not severely
compromised by blending, the line is fitted with a Gaussian profile,
and its central wavelength ($\lambda_{0}$), equivalent width
($W_{\lambda}$), and full width at half maximum (FWHM) are stored in a
table.  The uncertainty in $W_{\lambda}$ is calculated, taking
into account statistical errors in the data points and systematic
uncertainty in the placement of the local continuum. A line scan is
performed separately for each FUSE detector segment to avoid
introducing systematic errors that may be associated with combining the
overlapping segments into one dataset. Segment-by-segment analysis is
common practice among FUSE investigators and we adopt it here. After
each scan, lines that are blended or that were excluded from the scan
for some other reason are fitted separately and placed in the table.
There is one table for each of the four FUSE detector segments
(1a,1b,2a,2b) and two channels (LiF and SiC), for a total of eight
files.

At the conclusion of scanning, these tables contain the measurements
for all the suitable \h2\ lines in the spectrum, including multiple
measurements for those lines that appear on more than one detector
segment. There is a maximum of four measurements for each line.
Prior to COG fitting, these tables are merged with the following
scheme. The final $W_{\lambda}$ measurement for each line is the
average of the individual measurements, weighted by the inverse
squares of their individual uncertainties. The uncertainty in the
final measurement is the maximum of these three quantities: (1) the
quantity $(\sum \sigma _i^2)^{-\frac{1}{2}}$, where the $\sigma _i$
are the individual uncertainties; (2) one-half the difference between
the smallest and largest measurements (as a proxy for the standard
deviation); or (3) 10\% of the weighted mean itself.  For most lines,
(1) obtains the maximum value. However, for weak lines (2) is useful
if duplicate measurements are widely separated and the formal standard
deviation has little meaning. Finally, (3) is a conservative assumption
that attempts to account for unquantified or unknown systematic errors
in a uniform fashion.

After the individual line tables are combined, the software produces
an automated fit to a curve of growth with a single Doppler $b$
parameter. This simple assumption is maintained throughout and is
never clearly violated. Tests performed on high-quality data could
not confirm the need for multi-valued $b$ increasing with J, an effect
seen in {\em Copernicus} data (Spitzer et al.~1974) and attributed to
unresolved components with different rotational excitation.  We use a
downhill-simplex method to minimize the reduced $\chi^{2}$ statistic,
with column densities N(J) and $b$ as parameters.  We typically obtain
reduced $\chi^{2}$ in the range 0.5 - 2.0, with lower values for higher
signal-to-noise data.

The simplex $\chi^{2}$-minimization technique is fully automated and
well suited to deriving best-fit column densities and $b$, but it
is inadequate for deriving realistic uncertainties in the measured
parameters. The usual scheme of using the covariance matrix at the
best-fit point to derive confidence intervals on the parameters is not
appropriate in the highly nonlinear, multi-dimensional parameter space
of the curve of growth. Typically, the quantity $\partial \chi^{2} /
\partial p_{i}$ varies substantially with parameter $p_{i}$, such its
value at the best-fit point does not accurately describe a midpoint
between the critical $\chi^{2}$ values corresponding to the 1 $\sigma$
confidence interval.  So, while we use a standard minimization technique
for finding the best-fit column densities and $b$, we take a different
approach to finding the maximum 1$ \sigma$ variations in the parameters
(the error bars in Tables 5 and 6).

In this multi-dimensional parameter space, the column densities N(J)
and the Doppler $b$ parameter have somewhat different status, with the
uncertainty in $b$ contributing most of the uncertainty in N(J).  We first
calculate the critical value, $\chi^{2}_{\rm crit}$, corresponding to the
1 $\sigma$ confidence interval. This quantity is given by the best-fit
$\chi^{2}$ plus an offset $\Delta\chi^{2}$, which is distributed for
$N$ parameters like $\chi^{2}$ for N degrees of freedom (Bevington \&
Robinson 1992).  Over a range of $b$-values containing the best fit,
the software produces a curve-of-growth fit over a succession of $b$
values, in 0.1 km s$^{-1}$ intervals. At each fixed $b$, the best fit is
obtained by allowing N(J) to vary freely. If this best fit has $\chi^{2} <
\chi^{2}_{\rm crit}$, then the column densities for different J are varied
randomly to explore the parameter space in the immediate region of the
best fit with fixed $b$. Randomly chosen column densities that satisfy
the $\chi^{2}$ criterion are stored in a table for further analysis.
This process is repeated over a range of fixed $b$ values, building a
table of permitted parameters that satisfy the $\chi^{2}$ criterion.
The quoted uncertainties in the published N(J) and $b$ represent the
maximum variation among all the parameter sets that have $\chi^{2}
< \chi^{2}_{\rm crit}$, and they represent the true 1 $\sigma$ joint
confidence interval on the parameters. 

In addition to its efficiency, this technique has the further benefit
of providing realistic asymmetric error bars on the measurements.
This technique is more efficient than a deterministic multi-dimensional
grid, which requires many more COG evaluations if it is ``square'' in the
parameters. By contrast, our technique does not vary the column densities
if the best fit for a given $b$ is excluded. It requires approximately
10,000 COG evaluations for each sight line, far less than a deterministic
grid with sufficient resolution to capture all the variations in the
parameter space.

After COG fitting and exhaustive parameter space exploration, we have
derived N(J), $b$, and uncertainties for each line of sight.  However,
for diffuse clouds with N(\h2) $\gtrsim 10^{18}$ cm$^{-2}$, it is better
to derive N(0) and N(1) by fitting the wings of the damping profiles.
For lines on the damping part of the COG, we combine the COG fitting
technique described above with profile fitting. Model line profiles
are produced with parameters N(J) and $b$ for each \h2\ component
in the spectrum, and convolved by a Gaussian instrumental line spread
function. Typically, there are two components, corresponding to Galactic
and Magellanic gas.  Fitting is performed over the range of wavelength
necessary to capture all the relevant lines and exclude difficult regions
of the stellar continuum.  The local continuum in the region of the fitted
line profiles is described by a quadratic function.  We also add atomic
and ionic lines where necessary to improve the fit. Although \h2\ lines
with J $\geq$ 2 are included in the fitting, their final column densities
are instead calculated using the COG fitting technique.  Finally, we fit
the Lyman bands separately on each FUSE detector segment and combine the
measurements in a weighted average for final results. The uncertainties
quoted in Tables 3 - 6 are the standard deviations of the multiple
measurements (a more thorough discussion of this fitting is given by
Rachford et al.~2001). We find remarkable consistency among the different
Lyman bands and the different detector segments, giving us confidence
that 10\% uncertainty can be achieved routinely with this fitting method.

In a few cases, we derive very small uncertainties on N(0) and N(1).
The most extreme case is Sk -69 246 for which the uncertainties are
less than 10\%.  To test whether the data quality supports such small
uncertainties, we have performed a Monte Carlo simulation of the
effects of noise on the fits.  We generate a noiseless profile based on
the observed column densities, and then add a series of random noise
vectors whose characteristics match the observed noise in the object
spectra.  For Sk -69 246 we generated 10 noisy synthetic spectra and
performed profile fits.  The standard deviations of the values of N(0)
and N(1) were 0.01 dex, smaller than the reported uncertainties. This 
result supports the claim that the segment-to-segment variations in 
the fitted values dominate the total uncertainty. 

In our line analysis, COG fitting, and profile fitting, we have
paid careful attention to statistical and systematic uncertainties
in deriving \h2\ column densities and Doppler parameters.  The major
source of uncertainty in the column densities derived from COG fitting
is the often weak constraint on the Doppler $b$ parameter.  Generally,
lines from a given rotational level with a wide range of line strengths
($f\lambda$, the oscillator strength times the rest wavelength in \AA)
are necessary to adequately constrain $b$ for a given sight line (see
Figure~\ref{flfig}).  In practice, the best constraints are provided by
the relatively weak Lyman 1-0 and 0-0 bands.  If lines with both high and
low $f\lambda$ are available, or for lines on the linear portion of the
COG, the logarithmic errors in the column densities can approach 0.10
dex, and vary with the signal-to-noise ratio. If the available lines
for a given J-level are spaced narrowly in $f\lambda$, the error in
this column density will be determined largely by the permissible range
in $b$.  In this case, solutions for the N(J) are coupled together by
the $b$ value; lowering or raising $b$ will move all the N(J) up or down
together, respectively. Large uncertainties in $b$ can also cause very
asymmetric error bars on the column densities. If the best-fit column
density for a given $b$ lies near the best-fit N(J) for the extreme end
of the $b$ range, then the N(J) may take on a large uncertainty in the
other direction, corresponding to the other extreme of $b$.  All these
systematic variations make the ratios in the column densities (seen below
in \S~3) less uncertain than the individual column densities themselves.

A further systematic uncertainty is due to the unknown component structure
of the \h2\ clouds. At the 15 - 30 km s$^{-1}$ resolution of FUSE,
we generally cannot resolve the fine-scale structure of complicated
interstellar components observed in atomic lines with higher resolution
(e.g., by Welty et al. 1997, 1999). Our analysis assumes that all the
\h2 resides in one component, and that absorption from all rotational
levels can be described by a single Doppler $b$. Snow et al.~(2000) and
Rachford et al.~(2001) used the observed Na~I component structure along
Galactic translucent-cloud sight lines to construct a multicomponent
curve of growth. Here, the stronger absorption makes it more likely that
\h2\ is contained in more than one component.  They found that, while the
measured column densities N(J) were sensitive to the component structure,
the differences in N(J) for lines on the Doppler (flat) portion of the COG
were small compared to the uncertainties associated with the flat
COG. Thus, we assume single-component structure and leave more detailed
analysis for future examination of individual sight lines from our survey.

The sight lines in which \h2\ is not detected receive a separate 
analysis. If there is no visible evidence of the strong Lyman 7-0 
R(0) and R(1) lines, we place upper limits on the equivalent 
widths of these lines and convert these limits to column density 
limits assuming a linear curve of growth. We use the following 
expression for the (4 $\sigma$) limiting equivalent width of an unresolved 
line at wavelength $\lambda_{0}$:
\begin{equation} 
W_{\lambda} = \frac{4\lambda_{0}}{(\lambda / \Delta \lambda)\,(S/N)}, 
\end{equation} 
where $\lambda / \Delta \lambda$ is the spectral resolution, and S/N
refers to the signal-to-noise ratio {\em per resolution element} in the
1-2 \AA\ region surrounding the expected line position. By contrast,
in Tables 1 and 2 we list the S/N ratio per pixel to provide a means
of comparing datasets in a resolution-independent fashion.  Because the
resolution of FUSE varies across the band and even between observations,
we conservatively assume $\lambda / \Delta \lambda$ = 10,000 for all
upper limits. The N(0) and N(1) limits in Tables 5 and 6 are imposed
with the strongest J~=~0 and J~=~1 Lyman lines, 7-0 R(0) and 7-0 R(1).
For a linear COG, these lines have equivalent widths $W_{\lambda} =
(26.8$ m\AA) (N(0) / 10$^{14}$ cm$^{-2}$) and $W_{\lambda} = (18.7$
m\AA) (N(1) / 10$^{14}$ cm$^{-2}$), respectively. The entries 
in Tables 3 and 4 show a limit on the sum of the two individual lines.
Most of the upper limits lie just below the lowest detections of \h2.

In summary, we describe the results and uncertainties that appear in
Tables 5 and 6. We list the final column density and uncertainty results
for all our program stars. These column densities were derived from the
COG or profile-fitting routines described above. The uncertainties were
calculated from either COG parameter-space searches or band-to-band and
segment-to-segment variations in the profile fits, again as described
above. The 4 $\sigma$ upper limits are derived according to Equation
(1) from individual signal-to-noise ratios of the data in the region
of of the Lyman 7-0 R(0) and R(1) lines. For these limits we assume an
effective resolution $R$ = 10,000. For low-column density sight lines
where no constraint on the Doppler $b$ parameter is provided, we assume
that the lines fall on the linear portion of the curve of growth and label
these cases accordingly. These tables form the fundamental database for
all subsequent analysis.

\subsection{Notes on Individual Sight Lines} 

The LMC targets Sk -69 243 (R136) and MK 42 lie at the center of the 30
Doradus H~II region, and Sk -69 246 lies in an isolated field 4$'$ to the
north.  The Sk -69 246 spectrum shows no evidence of unusual properties.
The other two datasets, however, show line profiles much broader than the
expected FUSE line spread function in both the LMC and Galactic absorption
lines due to crowding among the 70+ hot stars in the inner 30$''$ of
the cluster.  Because the blurring mimics a low-resolution spectrograph,
we can only be sure that there is \h2\ at LMC velocities in these sight
lines, with N(\h2) $\simeq 10^{19}$ cm$^{-2}$.  The crowded fields and
line broadening preclude a detailed analysis of the column densities. We
include these stars in the \h2\ detection statistics (Tables 1 and 3)
but exclude them from the LMC column density sample (Table 5).

The sight lines to NGC 346 (3, 4 and 6) lie within a 50$''$ circle on the
sky in the SMC.  As seen in Tables 4 and 6, their column densities and
Doppler $b$ parameters are statistically indistinguishable, suggesting
that these closely spaced sight lines probe the same interstellar
cloud. In the abundance and excitation distributions and tests performed
below, we use the mean column density in each level and the mean $b$
to stand in for these three sight lines.

Mallouris et al. (2001) reported tentative detections of J = 1 and J =
3 lines in towards the SMC star Sk 108. Our survey imposes formal 4
$\sigma$ limits on the equivalent widths of undetected \h2\ lines.
According to this criterion, these lines are not significant, and we
place an upper limit on N(\h2) for Sk 108 that is larger than the value
they report.

For closely paired sight lines we can use the observed \h2\ excitation
to estimate the sizes of diffuse interstellar clouds.  The sight lines
to HD 5980 and AV 232 are separated by 58$''$ on the sky, and lie
$\simeq$ 17 pc apart at the distance of the SMC. Shull et al. (2000)
reported different rotational excitation N(J) in these two sight lines,
indicating that they may probe different molecular gas.  The Galactic
components in these sight lines show similar column densities and kinetic
temperatures, and are likely to arise from the same cloud.  However, the
SMC \h2\ components on these two sight lines show distinct differences
in rotational excitation.  Measurement of \h2\ excitation and abundance
in multiply-intersected absorbers will allow us to constrain the sizes
and structure of diffuse interstellar clouds.

\subsection{Ancillary Data} 

To fully exploit the FUSE data on \h2, we require additional
information about the dust and gas along the sight lines.  The most
important of the supplemental data is the atomic hydrogen column
density along each sight line.  To obtain N(H~I) for our entire 
sample, we use 21 cm radio emission and apply a correction to 
account for known systematic errors. 

We obtain H~I column densities from the recent Parkes and Australia
Telescope Compact Array maps (SMC, Stanimirovic et al.~1999; LMC,
Staveley-Smith et al.~2001) and then apply a correction, discussed
below.  These maps were created with $60''$ (LMC) and $98''$ (SMC)
beams and were provided to us in fully calibrated, machine-readable
form (L.  Staveley-Smith, personal communication).  To obtain the
N(H~I) for the FUSE sight lines, we select the four beam positions that
surround each star and perform a bilinear interpolation.  Using N(H~I)
from 21 cm emission may be subject to systematic uncertainties
associated with dilution in the 1$'$ beam if the interstellar gas is
highly clumped on smaller scales or if there is additional emission
behind the stars. Available UV absorption measurements provide a valuable
check on these measurements.

Attempts to derive N(H~I) from the Ly$\beta$ lines in the FUSE band
generally fail, owing to the severe blending between the Galactic and
SMC/LMC components, geocoronal O I emission, and the presence of the
\h2\ absorption. The higher Lyman series is present in the FUSE band
but is severely compromised by the dense forest of \h2\ and atomic
lines below 1000 \AA.  Data on H~I Ly$\alpha$ are available for roughly
two-thirds of our sample, but we find that reliable determinations of
N(H I) from a combination of Ly$\alpha$ and Ly$\beta$ profile fitting
is impractical in all but the best cases.  The largest uncertainties
are in the placement of the stellar continuum near these lines and in
the close velocity spacing of the Galactic and Magellanic components
(100 - 300 \kms).  While determinations could be made for roughly 30\%
of the targets with high-resolution, high-S/N data (from HST/FOS, GHRS,
and particularly STIS) and simple continuum structure, we could not
achieve this goal for the entire sample. We performed simple fits to
the Ly$\alpha$ profiles to estimate N(H~I) for the one-third of the
sample where the profile is not severely compromised.  In these trials,
we found that the constraints on N(H~I) range from 0.3 to 1.0 times the
21 cm value, with a mean of 0.56 for both Clouds.  We have not attempted
to remove contaminating absorption and complicated features in the
stellar continuum, which are present even in these best cases.  Thus,
fitting the Ly$\alpha$ profile is itself uncertain, and generally
establishes an upper limit. The lower limit is more difficult to
determine because the Magellanic component can be reduced and the
Galactic component increased to compensate.  However, it appears that
for 25 sight lines, the upper limit derived from Ly$\alpha$ is, on
average, half its value from 21 cm emission. This result suggests that
we should adjust our H~I column densities to account for this
systematic error.

We are reluctant to apply a uniform factor of 0.5 correction to the
entire FUSE sample, because the subsample may have internal biases. For
example, the stars with HST data may be located preferentially on the
front sides of the Clouds.  As a compromise between uniformity and
uncertainty, we reduce the 21 cm N(H~I) by a factor 0.75.  We then
impose a large uncertainty on N(H~I) to account for this systematic
error. For the final column density, we adopt N(H~I) =
($0.75^{+0.25}_{-0.25})\,\, \times$ N(21 cm).  This scheme captures the
uncertainty in the column density and can be applied to the entire
sample. Both extremes, at the high end from 21 cm data and at the low
end from our HST estimates, are included in the error bars.  The final
uncertainty in the molecular fraction is dominated by the uncertainty
in N(H~I).


The color excess, E(B-V), is an important measure of the dust abundance
towards the target stars.  The color excesses were obtained using
two methods:  (1) the observed B-V color and an adopted intrinsic
(B-V)$_{0}$ scale (Fitzgerald 1970 and Schmidt-Kaler 1982, with
-0.31 or -0.32 adopted for early giants and supergiants), and (2) the
observed UV spectrophotometry (IUE) plus B and V together with model
atmospheres to fit the spectral energy distribution.  In practice, we
take the mean of the two methods to minimize any external uncertainties.
Because intrinsic colors for Wolf-Rayet (WR) stars vary dramatically,
model atmospheres together with UV and optical spectrophotometry are
necessary to derive E(B-V).  We estimate the final uncertainty in these
values to be $\leq$ 0.02 mag.  A further uncertainty is the contribution
of foreground dust to the total sight line reddening.  We derive the
corrected color excess, E$'$(B-V), by subtracting from the measured
color excess a foreground E(B-V) = 0.075 for the LMC and 0.037 for the
SMC. These average values were obtained by the COBE/DIRBE project by
averaging the dust emission in annuli surrounding the Clouds (Schlegel,
Finkbeiner, \& Davis 1998).  In \S~3 we describe a test of these averages
with the observed gas-to-dust ratio.

We use the {\em Copernicus} survey of Savage et al.~(1977) to compare the
similarities and differences between the FUSE sample and \h2\ in the disk
of the Galaxy.  The sample of 109 stars obtained by S77 (their Table 1)
was culled for the 51 stars with measured N(H~I), either a  measurement
or limit on \h2, and E(B-V) $\leq 0.20$. The 17 upper limits in this
sample are included in certain tests depending on the context of the
comparison. In addition, we compiled column densities for $J \geq 2$
for 31 of these stars from Spitzer, Cochran, \& Hirshfeld (1974). These
column densities are compared to the rotational excitation of the LMC
and SMC gas in \S~3.

\subsection{Sample Selection and Bias} 

Most of the FUSE sample of LMC and SMC stars was not chosen specifically
for the \h2\ survey.  Most of the sample is drawn from three large FUSE
team programs devoted to stellar and interstellar research. The stellar
wind (Program P117) targets were chosen to occupy a range of spectral
types. The allotted time for each observation was 4 ksec, so this is
essentially a flux-limited sample.  In practice, this criterion selects
against stars with substantial extinction, which would be favored in a
sample chosen to study \h2.  The stars in Program P103, the FUSE study of
hot gas in the Magellanic Clouds, were chosen based on the availability
of supporting observations and to probe the supershells and bubbles
in the ISM of the LMC. In effect, this program selected against denser
regions and high extinction in favor of regions where cavities of hot gas
are expected.  In the SMC, seven stars were chosen specifically for \h2\
studies (Program P115).  All others were drawn from the hot star or hot
gas programs.  However, from the point of view of this survey, the stars
are located randomly across the LMC and SMC, but probably preferentially
situated on the Milky Way side of the dense absorbing regions in the
two galaxies.

\begin{figure*}[t] 
{\centerline{\epsfxsize=\hsize{\epsfbox{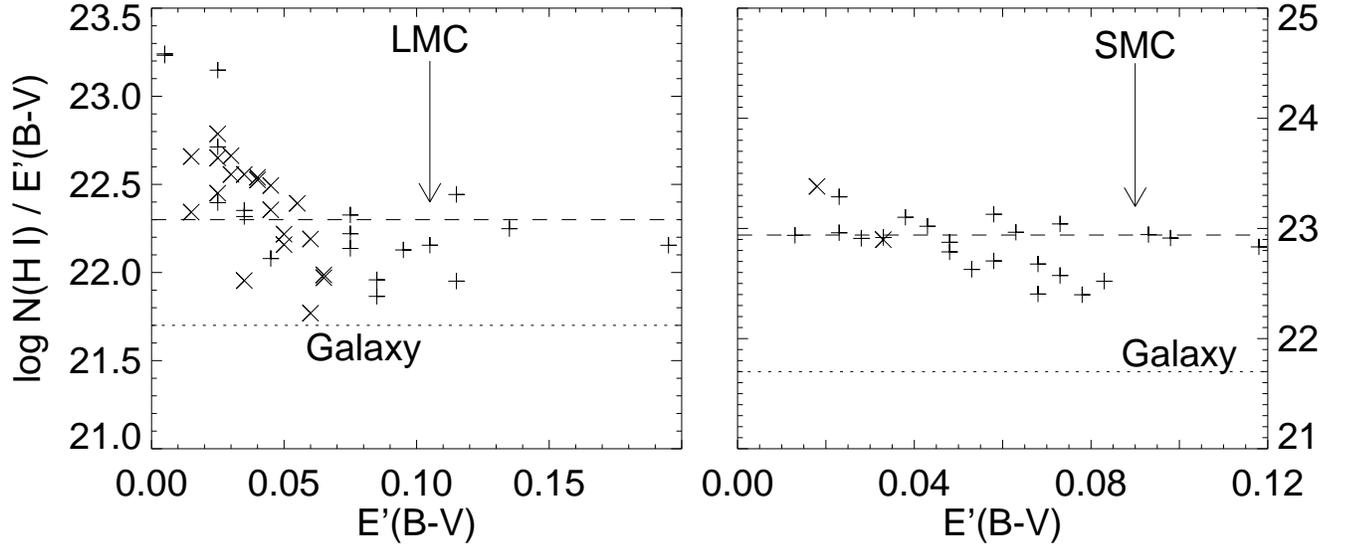}}}
\vspace{0.2in} 
\figcaption{\scriptsize Comparison of the gas-to-dust ratio, N(H~I) / E$'$(B-V), where
E$'$(B-V) for the LMC (left) and SMC (right) subtracts off the Galactic
contributions.  Detections of \h2\ are marked with $+$ and non-detections
are marked with $\times$.  The average ratio for the Galactic disk from
Shull \& van Steenberg (1985) is marked with a dotted line.  The average
LMC value (Koornneef 1982) and the average SMC value (Fitzpatrick 1985)
are marked with dashed lines. This concordance indicates that these sight
lines may sample most of the diffuse neutral gas detected by the 21 cm
emission beams, making the observed reduced molecular abundances robust.
\label{gas_to_dust}} \vspace{0.2in}} \end{figure*}

In addition to these selection biases, there is the additional problem
that flux-limited or extinction-limited surveys favor targets on the
foreground face of the LMC and SMC. This bias complicates the use of
H~I column densities derived from emission measurements. The 21 cm
emission beam samples all the gas, including some that lies behind the
target star but does not appear in absorption along the FUSE
pencil-beam sight lines.  In this case, the measured gas-to-dust ratio,
or N(H~I) / E$'$(B-V), will be enhanced over the average ratio
determined using H~I Lyman series absorption to derive N(H~I).
Figure~\ref{gas_to_dust} compares the corrected color excesses for the
program stars with their H~I column densities from 21 cm emission,
corrected as in \S~2.4.  The values of N(H~I)~/~E$'$(B-V) range from
$10^{22} - 10^{23}$ cm$^{-2}$ mag$^{-1}$ for the LMC stars, with a
general trend towards decreasing values of the ratio with higher
E$'$(B-V). There is a similar trend in the SMC, with N(H~I)~/~E$'$(B-V)
= $10^{22.5} - 10^{23.5}$ cm$^{-2}$ mag$^{-1}$.  We plot the average
value of this ratio for targets in the disk of the Milky Way as
determined from IUE measurements by Shull \& van Steenberg~(1985), and
the average values for the LMC (Koornneef~1982) and the SMC
(Fitzpatrick~1985). Our target stars lie well above the Galactic
average, but are consistent with the LMC and SMC values. Because
E$'$(B-V) appears in both axes of this plot, the error bars on the
points should run diagonally from the upper left to the lower right of
each point.  This fact probably explains the trend seen in the points,
with lower E(B-V) lying above the average and higher E(B-V) below. We
conclude that, while the location of the stars in the gas layers of the
LMC and SMC may account for some of the scatter in the molecular
fraction results, there is no large systematic increase in N(H~I) and
no corresponding systematic change in quantities derived from it. 
The observed trends suggest that our adopted correction to the 21
cm H~I columns is reasonable.  Had we adopted a larger correction to
the 21 cm columns, the points that lie below the average gas-to-dust
ratios would be far more discrepant.  For further discussions of
uncertainty in N(H~I), see \S~2.4. 

\section{SURVEY RESULTS AND ANALYSIS}

\subsection{General Results} 

A summary of results appears in Tables 3 and 4, where we list detected
column densities N(\h2), or upper limits, for the program stars.
Tables 5 and 6 list individual rotational level column densities N(J)
for the LMC and SMC, respectively, as derived from the curve-of-growth
and/or line-profile fits (\S~2).

The immediate, striking result of this survey, apparent from Tables
3-6, is the lower frequency of \h2\ detection in the LMC compared with
the SMC.  Shull et al.~(2000) reported seven sight lines through the
Galactic disk and halo, only one of which did not show detectable \h2 (PKS
2155-304). The detection rate in the initial FUSE sample of extragalactic
targets exceeds 90\%.  However, the total rate of detection in the LMC
is 23 detections in 44 sight lines, for a 52\% success rate. In the SMC,
the detection rate is 24 of 26, or 92\%. This difference appears to
be related to the patchy structure of the LMC ISM, which contains many
identified H~I superbubbles blown out by OB associations and supernova
remnants (Kim et al.~1999). By comparison, the ISM of the SMC is more
quiescent and less patchy (see \S~4-5 for further discussion).

\subsection{\h2 Abundance} 

The lower abundance of \h2\ in the low-metallicity gas of the LMC and SMC
is a key result of this survey. The molecular hydrogen abundance is sensitive
to the formation/destruction equilibrium of \h2\ in diffuse gas, where
far-ultraviolet (FUV) radiation competes with formation on dust grains to
determine the relative abundance of \h2.  The size of the sample allows
us to apply statistical tests of \h2\ abundance in the LMC and SMC
and compare it to H~I, dust properties, and diagnostics of
the UV radiation field.  The reduced molecular fraction may indicate
less efficient formation of \h2\ molecules on interstellar dust grains
or enhanced \h2\ photo-dissociation by an intense FUV radiation field,
or a combination of these two effects.  In this section, we use simple
analytic expressions and numerical models to assess the effects of varying
environmental conditions on the \h2\ formation/destruction equilibrium.

\begin{figure*}[t] 
\centerline{\epsfxsize=\hsize{\epsfbox{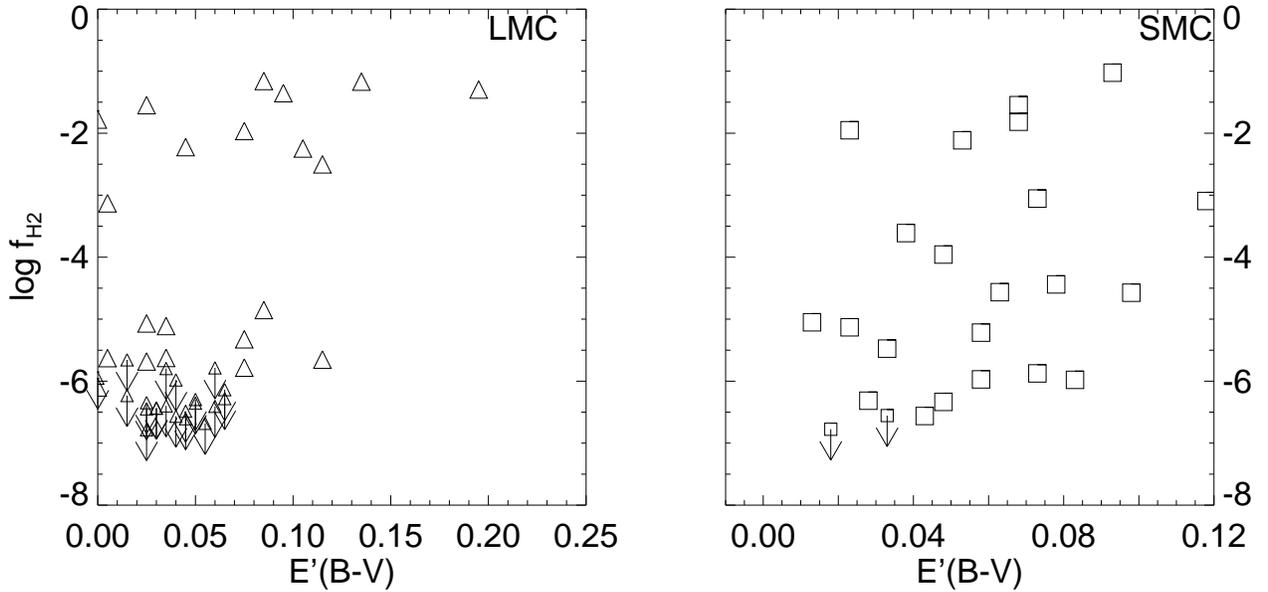}}} 
\figcaption{\scriptsize Molecular fraction, \fh2\ = 2N(\h2) / [N(H~I) + 2N(\h2)],
versus corrected color excess. E$'$(B-V) was determined from FUV
extinction measurements and is corrected for a mean Galactic E(B-V)
= 0.075 for the LMC and E(B-V) = 0.037 for the SMC.  For stars with
E$'$(B-V) $<$ 0.075 or 0.037, we have set the corrected value to E(B-V)
= 0.00. See the discussion in \S~3.
\label{molec_frac}} 
\end{figure*}

\begin{figure*}[h] 
\centerline{\epsfxsize=\hsize{\epsfbox{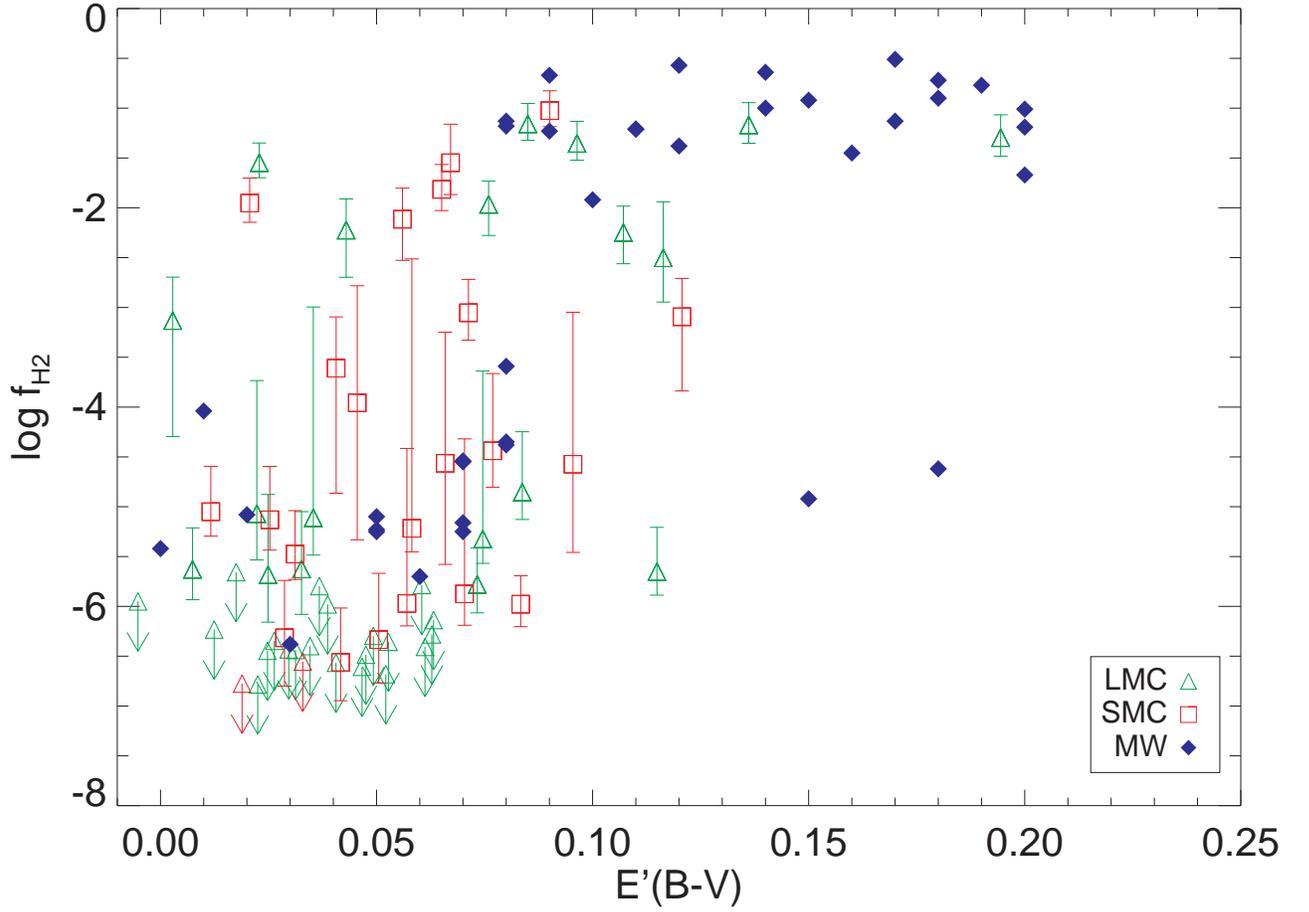}}}
\figcaption{\scriptsize Comparison of the FUSE LMC and SMC molecular fractions with
the {\em Copernicus} (S77) sample of stars in the Galactic disk (MW).  The SMC
and LMC points have been given small random shifts in foreground-corrected
color excess of up to E$'$(B-V) = $\pm$0.003 to separate coincident
points.  Below E$'$(B-V) = 0.08, the scatter in molecular fraction
is large in the Clouds. For E$'$(B-V) $\geq 0.10$, the SMC and LMC points lie
below the average Galactic value of \fh2 $\simeq 0.1$.
\label{corr_s77}}
\end{figure*}

To assess the total abundance of \h2, we define the quantity \fh2 =
2N(\h2) / [N(H~I) + 2N(\h2)], which expresses the fraction of hydrogen
nuclei bound into \h2. The molecular fractions for our sight lines appear
in Figure~\ref{molec_frac}, correlated with the corrected color excesses
of the target stars.  In Figure~\ref{corr_s77} we compare the S77 sample
of Galactic disk stars with the LMC and SMC targets.  The key difference
between the samples is the presence of several LMC and SMC sight lines
with \fh2 $\geq 10^{-4}$ at low E$'$(B-V).  In the Clouds we find a wide
range of molecular fractions at low reddening, E$'$(B-V) $\leq 0.08$.
In the LMC there is a noticeable gap near $\log$ \fh2\ $= -4$, which
does not appear in the SMC. The most striking pattern in these plots is
the large number of sight lines to the LMC that show no detectable \h2.
S77 found low values of \fh2\ at low E(B-V) and a transition from low to
high values (\fh2\ $\sim$ 0.1) at a color excess E(B-V) = 0.08.  This pattern
is not readily discernible in the samples.

Our results reverse prior indications that the average molecular fraction
in the LMC is not reduced relative to the Galaxy.  Using HUT, Gunderson et
al.~(1998) measured N(\h2) $\simeq$ 10$^{20}$ cm$^{-2}$ for two LMC stars,
Sk -66 19 and Sk -69 270, that do not appear in our sample. These sight
lines have \fh2\ = 0.04, similar to the highest-\fh2\ points in the FUSE
sample. They concluded that, in the low-metallicity LMC, N(\h2)/E(B-V)
did not differ from the average Galactic value, despite large differences
in N(H~I)/E(B-V).  With the larger FUSE sample, we find a substantial
reduction in \fh2\ in the LMC.  With ORFEUS data on three LMC stars,
Richter~(2000) found that, although N(\h2) relative to E$'$(B-V) in the
Clouds might be similar to the Galactic values, the N(\h2)/N(H~I) ratio
was lower than the Galactic average. The larger FUSE sample confirms
this result.

To quantitatively compare the distributions of \fh2\ for the Galaxy and
Magellanic Clouds, we performed Kolmogorov-Smirnov (KS) tests on the
samples.  The two-sided KS test finds a 0.224 probability that the S77
and SMC samples are drawn from the same parent distribution, and a
$<$0.001 probability for the S77 and LMC samples.  Compared to each
other, the SMC and LMC samples give a 0.121 probability of being drawn
from the same distribution.  This difference reflects the very
different detection rates in the Clouds, but they suggest that the
reduced \fh2\ in both the LMC and SMC is significant.

For a quantitative comparison of the Galactic disk and LMC molecular
fractions, we define the average column-density weighted molecular
fraction $\langle f \rangle$ to be:
\begin{equation} 
\langle f \rangle = \sum \rm 2 N(H_2) / \sum [N(\rm H~I) + 2 N(\rm H_2)]. 
\end{equation} 
In evaluating the average molecular fractions, we include the FUSE
upper limits as measurements, so that $\langle f \rangle$ from FUSE is
also an upper limit.  For the LMC, we find $\langle f \rangle = 0.012$
for stars with E$'$(B-V) = 0.00 - 0.20. For the SMC, we find $\langle f
\rangle = 0.010$ for stars with E$'$(B-V) = 0.00 - 0.12. For the
Galactic comparison sample, we find $\langle f \rangle = 0.039$ for
stars with \ebv\ = 0.00 - 0.12, for comparison with the SMC.  For
Galactic stars with \ebv\ = 0.00 - 0.20, we find $\langle f \rangle =
0.095$, for comparison with the LMC. If we exclude the upper limits in
Tables 5 and 6 from the sample, we find $\langle f \rangle = 0.010$ for
the SMC and $\langle f \rangle = 0.021$ for the LMC, due to the large
large number of non-detections there.

\begin{figure*}[t] 
\plotone{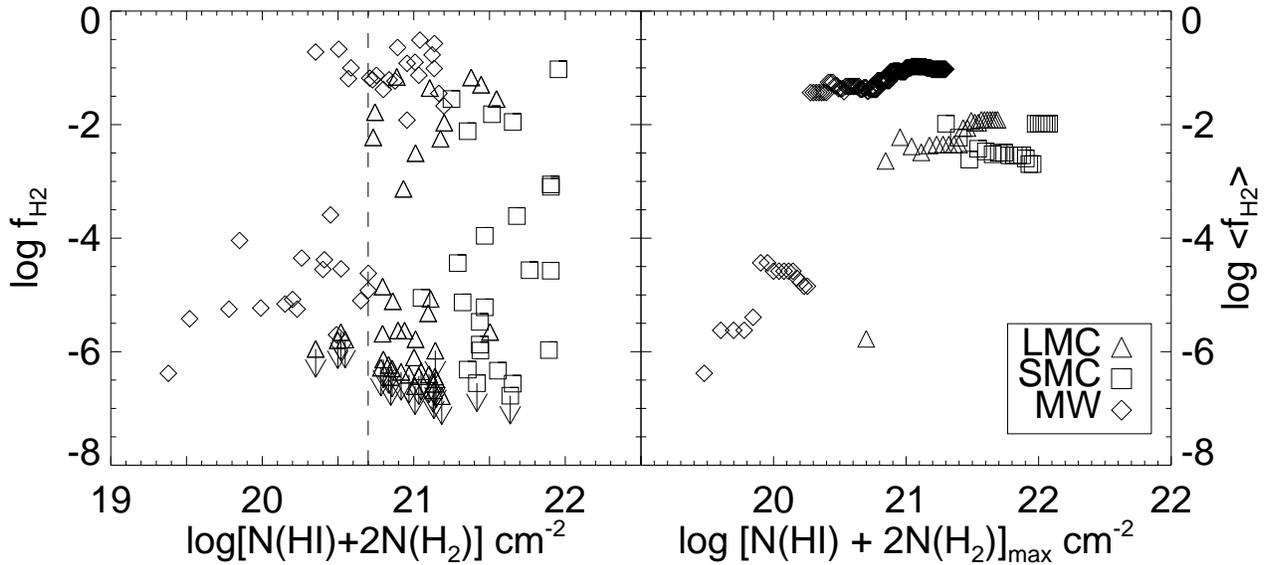}
\figcaption{
\scriptsize Left panel: Molecular fraction versus total H   
column density for the 44 LMC and 26 SMC sight lines and the S77 sample.
Right panel: Cumulative average molecular fraction for the LMC,
SMC, and S77 samples, as a function of the maximum value of the total
gas content, N(H) = N(H~I) + 2N(\h2), considered in the average. The
LMC and SMC are more gas-rich than the Galaxy, but they have a lower
molecular fraction.  
\label{compare_frac}}
\end{figure*}

Figure~\ref{compare_frac} shows the individual and cumulative molecular
fractions correlated with the total H content, N(H) = N(H~I) + 2N(\h2).
The S77 sample shows a clear break near log N(H) = 20.7, above which
all clouds have molecular fraction \fh2\ $\sim 0.1$.  This break is
usually identified as the point at which the cloud achieves complete
self-shielding from interstellar radiation (S77).  The LMC and SMC sight
lines show no such clear transition, even at the higher N(H) they probe.
As we show below, the LMC and SMC show evidence of reduced formation
rate of \h2\ on dust grains and of enhanced photo-dissociating radiation
relative to the Galaxy. In this case, we expect to see a transition from
small \fh2\ with large scatter to a tighter band of high \fh2, but at
a higher total H column density than in the Galactic disk.  The FUSE
samples require that such a transition occurs at $\log N(H) \geq 21.3$
in the LMC and $\log N(H) \geq 22$ in the SMC. Planned observations
of several LMC and SMC stars with E$'$(B-V) $\geq$ 0.20 will address this
point in the future.

The lower observed $\langle f \rangle$ is evidence for lower molecular
abundance in the LMC and SMC.  However, there is a potential systematic
effect that we must reemphasize.  The S77 survey used column densities,
N(H~I) and N(\h2), determined from the same {\em Copernicus} spectra.
This technique does not suffer the systematic uncertainty of comparing
measurements of N(H~I) from radio emission measurements with a large
beam and N(\h2) determined from pencil-beam sight lines. Thus, relative
to the S77 measurements, we may underestimate the molecular fraction
due to the systematic overestimate of the H~I column determined from 21
cm emission if a substantial fraction of the H~I seen in emission lies
behind the target star or if clumping is significant below 1$'$
scales.  However, because the average dust-to-gas ratio for our targets
scatter around the average values found for the LMC and SMC, this
problem is unlikely to exceed the generous error budget assigned to
N(H~I).  (See the discussion of these issues in \S~2).

A simple model of \h2 in an interstellar cloud (Jura 1975a,b) can
be expressed as an equilibrium between the formation of \h2\ with a rate
coefficient $R$ (in cm$^{3}$ s$^{-1}$) and the photodestruction of \h2\
with rate $D$ (in s$^{-1}$):
\begin{equation} 
Rn_{HI}n_{HI} = Dn(H_2) \simeq f_{\rm diss} \sum_{J}^{}\beta (J) n(H_2,J),
\end{equation} 
where $\beta (J)$ is the photoabsorption rate from rotational level J and
$n_H = n(H~I) + 2n(\rm H_2)$.  The coefficient $f_{\rm diss}$ corresponds
to the fraction of photoabsorptions that decay to the dissociating
continuum and varies from 0.10 to 0.15 depending on the shape of the UV
radiation spectrum.  Molecules are destroyed by photoabsorption in the
Lyman and Werner bands, followed by radiative decay to the vibrational
continuum of the ground electronic state.  The formation of \h2\ is
believed to occur on the surfaces of dust grains, but may also occur in
the gas phase at a lower rate (Jenkins \& Peimbert 1997).

If we further assume a homogeneous cloud and replace the number
densities $n_{\rm HI}$ and n(\h2) with column densities N(H~I) and N(\h2),
respectively, then:
\begin{equation} 
f_{\rm H2} = \frac{2 R n_{HI}}{D}, 
\end{equation} 
where \fh2\ is the molecular fraction defined above.  Thus, the observed
molecular fraction is a measure of interaction between formation and
destruction, and the molecular fraction can be suppressed both by reduced
formation rate on dust grains and by enhanced photodissociating radiation.

To assess the reduced molecular fraction, and, later, the rotational
excitation of the sight lines, we use the results of an extensive grid
of models produced by a new numerical code that solves the radiative
transfer equations in the \h2\ lines and produces model column
densities for a range of model parameters (Browning, Tumlinson, \&
Shull 2001, in preparation).  This code has been developed with the
FUSE surveys of \h2\ in mind.  The model clouds are assumed to be
one-dimensional and are illuminated on one side by a uniform radiation
field of specified mean intensity, $I$, in units photons cm$^{-2}$
s$^{-1}$ Hz$^{-1}$. (The code can illuminate the cloud on both sides,
but the resulting changes in the model column densities are small.)
generations of the code will permit two-sided clouds illuminated on
both sides.)  Because it follows the radiation field and \h2\ abundance
in detail, the code makes no distinction between the optically thin and
optically thick (self-shielded) regimes treated separately by Jura
(1975a,b).  The cloud density n$_{H}$, kinetic temperature $T_{\rm
kin}$, Doppler $b$ parameter (to approximate thermal or turbulent
broadening), cloud size $d$, and \h2\ formation rate coefficient on
grains ($R$, in units cm$^{3}$ s$^{-1}$) are the model parameters. The
comparison grid calculated for this work consists of 3780 models with
$I$ = (1.0, 4.0, 10, 20, 50, and 100) $\times\, 10^{-8}$ photons
cm$^{-2}$ s$^{-1}$ Hz$^{-1}$, $R$ = (0.1, 1.0, 3.0) $\times 10^{-17}$
cm$^{3}$ s$^{-1}$, n$_{H}$ = (5, 25, 50, 100, 200, 400, 800) cm$^{-3}$,
$d$ = (2, 4, 6, 8, 10) pc, and $T_{\rm kin}$ = (10, 30, 60, 90, 120,
150) K. For comparison with the LMC and SMC, we assume that
``Galactic'' values are $R = (1 - 3) \times 10^{-17}$ cm$^{3}$ s$^{-1}$
and $I = 10^{-8}$ photons cm$^{-2}$ s$^{-1}$ Hz$^{-1}$. In Table 7 we
define four models grids for comparison with the observed abundance and
excitation patterns.  These grids are chosen to have the full range of
$T_{\rm kin}$, $n$, and $d$ but vary in their assumed incident
radiation and \h2\ formation rate constants.

\begin{figure*}[t] 
\plotone{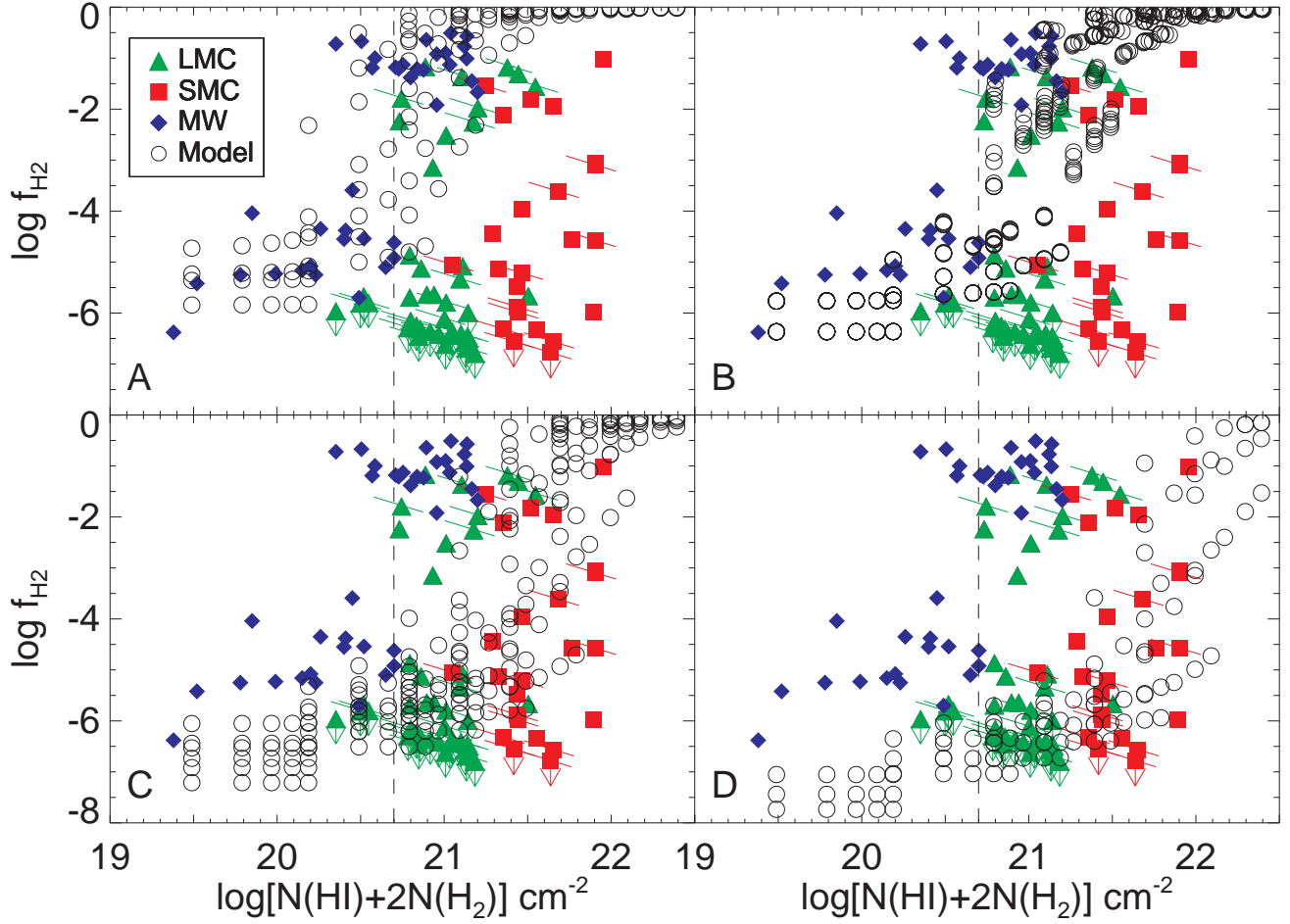}
\figcaption{\scriptsize The S77, LMC, and SMC samples plotted against
the total H content and compared to our numerical models. The panels
are labeled with the letter of the model grids they display (see Table
7). Panel A: These models have Galactic values of $R$, $I$, and the
full range of kinetic temperature, gas density, and cloud size. Panel
B: As before, but with $R = 3 \times 10^{-18}$ cm$^{3}$ s$^{-1}$, 0.1
times Galactic.  Panel C:  Same as before, with Galactic $R$ and $I =
10 - 100 \times$ the Galactic value.  Panel D:  As before, but with $R
= 3 \times 10^{-18}$ cm$^{3}$ s$^{-1}$ and $I = 10 - 100 \times$ the
Galactic value. The agreement with the LMC and SMC data indicates
enhanced radiation and reduced \h2\ formation rate in the Clouds.  See
\S~3.2 for more discussion. The error bars are attached randomly to
half the points to account for the systematic uncertainty in N(H~I). 
\label{f_model}}
\end{figure*}

In Figure~\ref{f_model} we display the LMC, SMC, and S77 samples together
with the model clouds, in terms of molecular fraction correlated with
total H content. First, we compare the data to model grid A, with typical
Galactic conditions (upper left panel).  These models match the Galactic
points reasonably well, but they do not coincide with the LMC and SMC
samples at all. The Milky Way agreement is a good starting point for
understanding the different conditions in the low-metallicity Clouds.

In Figure \ref{f_model} (upper right panel), we plot model grid B, with
a typical Galactic radiation field and a low formation rate constant,
$R = 3 \times 10^{-18}$ cm$^{3}$ s$^{-1}$.  This value is 3-10 times
below the Galactic values inferred from {\em Copernicus} data, $R =
(1 - 3) \times 10^{-17}$ cm$^{3}$ s$^{-1}$ (Jura 1974).  These model
clouds match some LMC points, but in general they are overabundant in
\h2, particularly at the column densities N(H~I) of the SMC sight lines.
In the lower left panel, we show the effect of enhancing the radiation
field incident on the model clouds to $I = 10^{-7} - 10^{-6}$ cm$^{-2}$
s$^{-1}$ Hz$^{-1}$, 10-100 times the Galactic mean (Jura 1974), while
fixing $R$ at the Galactic value (model grid C).  This adjustment again
produces the correct change in the pattern, reducing the molecular
fraction and coinciding with some LMC points.  Finally, the lower right
panel displays model grid D, raising the radiation field by factors of
10 - 100 and lowering R to 1/3 - 1/10 the Galactic value.  Only in this
extreme case are we able to produce an abundance pattern that resembles
the SMC and LMC samples. The observed clouds have relatively high N(H~I),
but lower average \fh2\ than the S77 sample. A combination of reduced $R$
and enhanced $I$, though not unique, can reproduce the observed patterns.

We have made no attempt to derive individual model parameters for each
of the sight lines in the sample. Such solutions are not unique in
the absence of independent constraints on the cloud density and kinetic
temperature (Browning, Tumlinson, \& Shull 2001, in preparation).  Indeed,
such sight-line modeling is unnecessary for a large sample that can be
compared to similar ensembles of model clouds.  The non-uniqueness problem
and the large variation in the observed conditions implied by the LMC and
SMC abundance distributions hinder us from deriving unique quantitative
constraints on $\beta$ and $R$ for the sight lines. However, we have
shown that realistic grids of model clouds that accurately reproduce
well-known Galactic patterns can match the LMC/SMC molecular abundance
patterns seen in the FUSE samples.  These models assume high incident
radiation fields and a low grain formation rate coefficient.

\subsection{\h2\ Excitation} 

We cannot determine from the abundance information alone whether the data
favor suppressed molecule formation in low-metallicity gas or enhanced
photo-dissociation by UV radiation as an explanation for the reduced
$\langle f \rangle$.  The most likely scenario is a combination of
these two effects.  To separately diagnose the effects of the competing
formation/destruction processes, we examine the rotational excitation
of the \h2.  

Following the absorption of a FUV photon and excitation to
an upper electronic state, the \h2\ molecule fluoresces to the ground
electronic state and cascades down through the rotational and vibrational
lines. The net effect of these repeated excitations and cascades is a
redistribution of the molecules into the excited rotational states of
the ground electronic and ground vibrational state. We observe this
distribution of molecules directly in our data, and derive column
densities N(J) in each rotational level of the ground vibrational state.

\begin{figure*}[t] 
\plotone{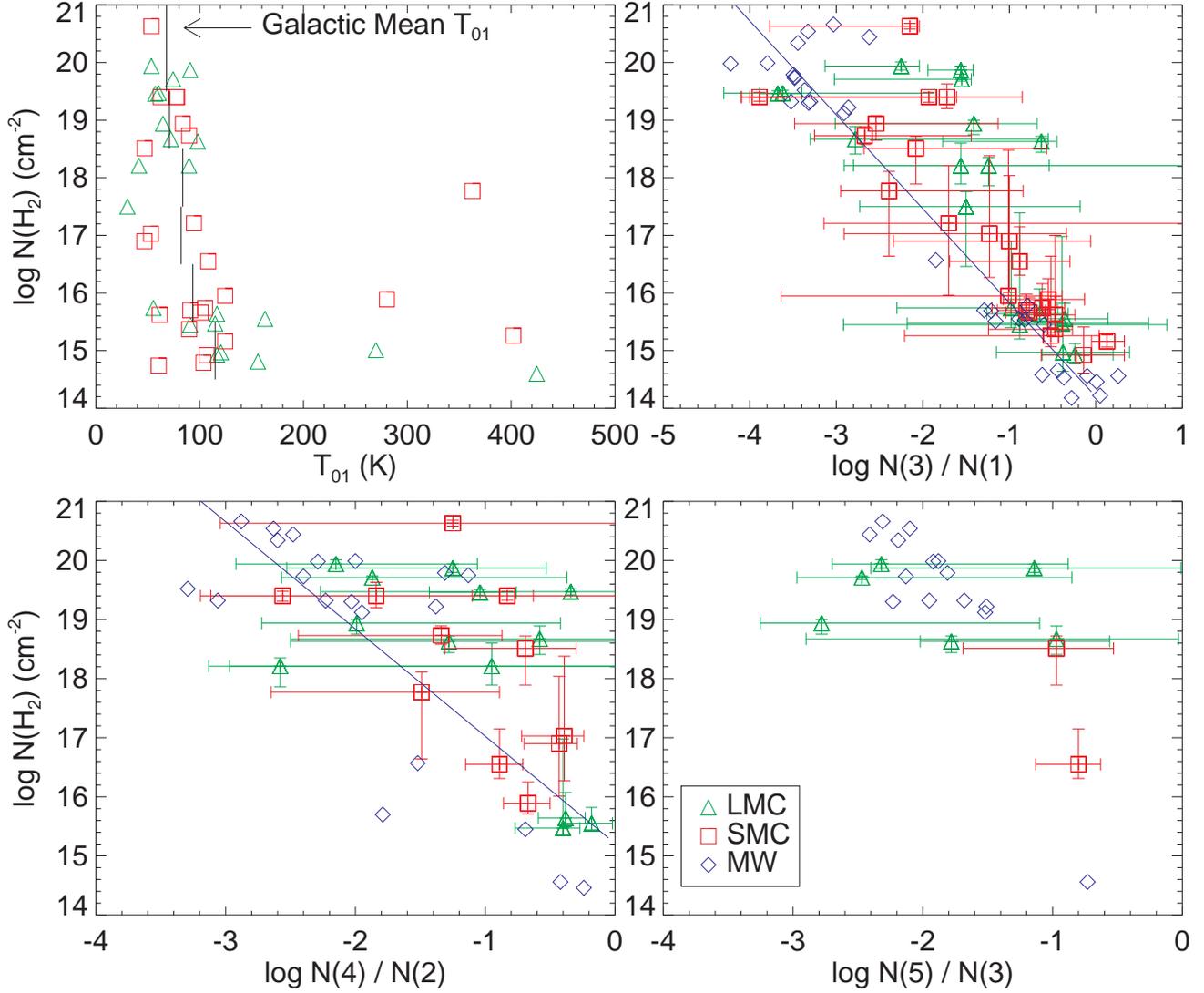} 
\figcaption{\scriptsize Upper left: Rotational temperature \t01\ plotted against 
total \h2\ column density N(\h2).  We find $\langle T_{01} \rangle = 82
\pm 21$ K for sight lines with N(\h2) $\geq 10^{16.5}$ cm$^{-2}$
(excluding the outlier AV 47 in the SMC) and $\langle T_{01} \rangle =
115$ K for all sight lines.  The rotational temperatures for LMC and
SMC sight lines lie near the Galactic average of $77 \pm 17$ K from
S77, indicating that the physical conditions in LMC and SMC gas may be
similar to those in the Galaxy if \t01\ traces the kinetic temperature
for high column density clouds.  Upper right: The column density ratio
N(3)/N(1) for the LMC and SMC samples, compared with the Galactic
sample. There is no evidence for enhanced N(3)/N(1) in the Clouds,
perhaps indicating that this ratio is determined by collisions at
densities similar to those in Galactic gas. The blue lines are fits to
the Galactic points.  Lower left: The column density ratio N(4)/N(2)
for the LMC and SMC samples, compared with the Galactic sample.  Lower
right:  The column density ratio N(5)/N(3) for the LMC and SMC
samples, compared with the Galactic sample.
\label{ex}}  
\end{figure*}

The rotational temperature \t01\ is given by: 
\begin{equation} 
T_{\rm 01} = \frac{\Delta E_{01} / k }{\ln [(g_1/g_0) N(0)/N(1)]}, 
\end{equation} 
where the column densities are those given in Tables 5 and 6, the ratio
$g_1/g_0$ of statistical weights of J~=~1 and J~=~0 is 9, and $\Delta
E_{01} / k =$ 171 K.  The quantity \t01\ is thought to trace the kinetic
temperature of the molecular gas at high densities where collisions
dominate the level populations (such as the translucent cloud regime,
see Snow et al. 2000). In lower-column diffuse clouds, the relationship
between \t01\ and $T_{\rm kin}$ is not known exactly.  As a simple
test for widely divergent physical conditions in the Magellanic Clouds
relative to the Galaxy, we derive the rotational temperature \t01\
for our targets (Tables 3 and 4). In Figure~\ref{ex} (top left panel)
we plot total \h2\ column density versus \t01. A clear pattern emerges,
with widely varying \t01\ at lower N(\h2), where clouds are not expected
to self-shield and densities may be too low for collisions with H$^{+}$
or H$^{+}_{3}$ to establish a thermal ortho-para ratio. For all sight
lines with N(\h2) $\geq 10^{16.5}$ cm$^{-2}$, we find $\langle T_{01}
\rangle  = 82 \pm 21$ K (the standard deviation is calculated excluding
the outlier AV 47 in the SMC).  For all sight lines, we find $\langle
T_{01} \rangle  = 115$ K.  Thus, for likely self-shielded clouds, the
rotational temperatures lie near the Galactic average, 77 $\pm$ 17 K
(S77), providing evidence that the distribution of kinetic temperatures
is similar to that in Galactic diffuse molecular gas.  While \t01\ may
closely track the kinetic temperature, the measured Doppler $b$ parameters
indicate significant non-thermal motions and do not correlate with \t01.

To further explore the rotational excitation of \h2\ in the samples, we
use the column density ratios N(3)/N(1), N(4)/N(2), and N(5)/N(3), in
order of decreasing sensitivity to collisional excitation and
increasing sensitivity to the radiative cascade.  We plot these ratios
for the LMC, SMC, and Galactic comparison samples in Figure~\ref{ex}.  The
trend toward lower excitation at higher N(\h2) is apparent in all the
ratios. In N(3)/N(1), there is no clear distinction between the three
samples, perhaps indicating the importance of collisions in determining
this ratio. However, it appears that both the LMC and SMC samples are
enhanced in N(4)/N(2) and N(5)/N(3), perhaps indicating the presence of
a FUV radiation field above the Galactic mean.  The large error bars on
these ratios are a direct result of the uncertainties in the Doppler
$b$ parameter. 

\begin{figure*}[t] 
{\centerline{\epsfxsize=\hsize{\epsfbox{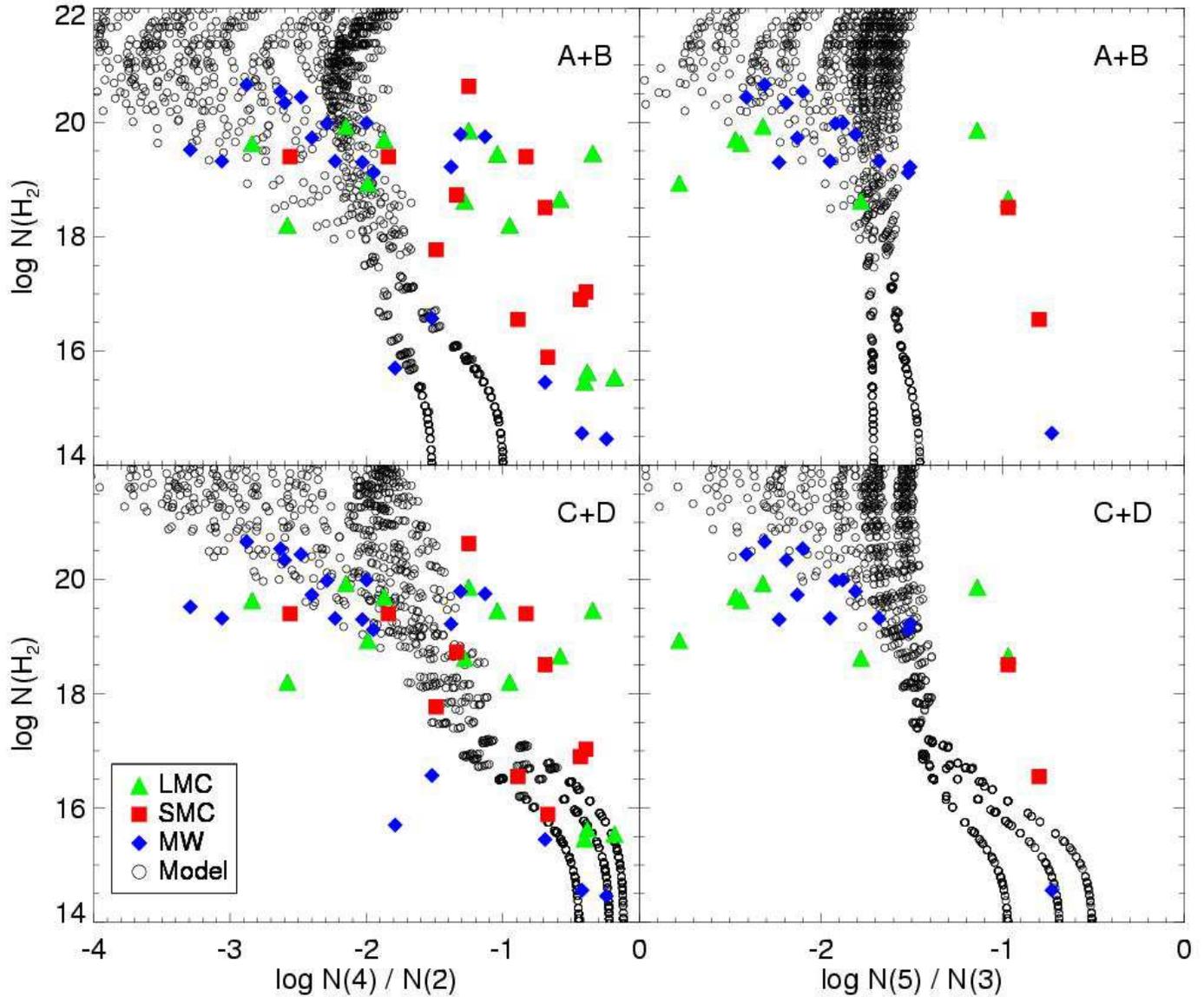}}}
\figcaption{\scriptsize The column density ratios N(4)/N(2) and N(5)/N(3) for
the LMC, SMC, and Galactic samples, compared with the model grid (in
open circles).  The model grids plotted have the range of density,
size, temperature, and incident radiation field discussed in \S~3.2. The
LMC and SMC samples show considerable rotational excitation above the
Galactic observations and the models.  In general, increasing incident
radiation and decreasing formation rate coefficient $R$ move model points to
the lower right on this diagram.
\label{ex_model}}}
\end{figure*}

In Figure~\ref{ex_model} we again plot the column density ratios N(4)/N(2)
and N(5)/N(3) compared with the model grids. In this comparison, the
difference between the Galactic and Magellanic samples is striking.
With the exception of six outliers, all the Galactic stars lie within
the region described by model grids A and B. In contrast, the majority
of the LMC and SMC points have N(4)/N(2) and N(5)/N(3) higher than the
model grid with Galactic conditions.

We consider four hypotheses to explain the unusually high excitation
ratios seen in the Clouds: (1) low kinetic temperatures elevate the ratios
by differentially enhancing J~=~0 over J~=~1, which are then redistributed
to J~=~4 and 5, respectively; (2) the upper levels are preferentially
populated in newly formed molecules (``formation pumping''); (3) high
incident FUV radiation pumps the upper levels via a radiative cascade;
(4) a low formation rate of \h2\ on grain surfaces allows FUV radiation
to penetrate to larger N(\h2), elevating the ratios in the cloud cores.
The results of our modeling follow these four mechanisms.

(1) Low kinetic temperature elevates the excitation ratios in model clouds
at fixed density, size, and incident radiation. The J = 4 and 5 levels
are more efficiently populated than J = 2 and 3  by absorptions from J =
0 and 1, followed by radiative cascade. Thus, any process that boosts
N(0) and N(1) increases N(4)/N(2) and N(5)/N(3).  The right envelope of
the models pictured in the upper left panel of Figure~\ref{ex_model}
is defined by $T_{\rm kin}$ = 10 K. Low $T_{\rm kin}$ by itself is
insufficient to explain the enhanced ratios.

(2) The idea that \h2\ molecules are formed on the surfaces of
interstellar dust grains has widespread theoretical and observational
support. However, the rotational and vibrational excitation of the \h2\
molecule as it exits the grain is unknown. Because of the high Einstein
$A$ values of the levels involved in the initial fluorescence, all memory
of the formation distribution is quickly erased in irradiated ensembles
of molecules. The current numerical code assumes that all molecules are
formed in a 3:1 ortho-para ratio in the highest vibrational level (v =
14) of the ground electronic state. This assumption was also made by the
models of van Dishoeck \& Black (1986). However, because in equilibrium
there is only one molecule formed for every 10 excitations by a UV photon
($f_{diss} \sim 0.1$), the formation distribution is quickly erased.
Indeed, test models produced by the current code over the range of
radiation inputs $I$ and assuming that half of all \h2\ is formed into J
= 4 and half into J = 5 do not differ appreciably from models made with
this general assumption. Thus, formation pumping is an unlikely cause
of the elevated excitation.

(3) As end points to the radiative cascade from the upper electronic
state, J~=~4 and J~=~5 levels are favored over J~=~2 and J~=~3.  Thus,
more frequent photoabsorptions result in higher N(4)/N(2) and N(5)/N(3)
in model clouds.  In Figure~\ref{ex_model} (lower panel) we show the
complete grid of models, which range up to $I = 1.0 \times 10^{-6}$
photons cm$^{-2}$ s$^{-1}$ Hz$^{-1}$, 100 times the Galactic mean (Jura
1974).  The points with $T_{\rm kin}$ = 10 K and this high radiation
field are still not consistent with the observed excitation ratios.

(4) A low grain formation rate of \h2 can elevate the excitation ratios
in model clouds. A smaller formation rate allows radiation to penetrate
to higher N(\h2) in deeper parts of the cloud, similar to the effects
of increasing the incident radiation field. However, models with $R =
3 \times 10^{-18}$ cm$^{3}$ s$^{-1}$, one-tenth the typical Galactic
value, do not reproduce the observed ratios.

Finally, we attempt to match the model clouds with the same set of
parameters that successfully match the \h2\ abundance patterns. The lower
panels of Figure~\ref{ex_model} show the grid of 1890 models with $R = 3
\times 10^{-18}$ and $I = 10^{-7} - 10^{-6}$. The combination of low $R$
and high $I$ produces the best match to the observed N(4)/N(2) ratios
and the N(5)/N(3) ratios at low N(\h2). Based on this concordance and
the match to the molecular fraction results above, we conclude that,
on average, diffuse \h2\ in the Magellanic Clouds exists in conditions
different from the Galaxy. These sight line models are not unique, but
both a high radiation field and a low formation rate coefficient are
necessary to explain the observed abundance and excitation patterns
in grids of model clouds that accurately reproduce the Galactic
distributions.

\begin{deluxetable}{lccc} 
\tablecolumns{4} 
\tablenum{7} 
\tablewidth{0pt} 
\tablecaption{\h2\ Model Cloud Grids} 
\tablehead{ 
\colhead{Label} 
&\colhead{$R (10^{-17})$} 
&\colhead{$I (10^{-8})$} 
&\colhead{Description} \\ 
\colhead{} 
&\colhead{cm$^{3}$ s$^{-1}$} 
&\colhead{ph cm$^{-2}$ s$^{-1}$ Hz$^{-1}$} 
&\colhead{}  
} 
\startdata 
A    & 1 - 3         & 1 - 4     &  Galactic conditions    \\ 
B    & 0.3           & 1 - 4     &  \\ 
C    & 1 - 3         & 10 - 100  &  \\ 
D    & 0.3           & 10 - 100  &  LMC, SMC \\ 
\enddata 
\end{deluxetable}

The model grids with high $I$ and low $R$ provide the best match to
the observed abundance and excitation of \h2, but there are still some
LMC and SMC points with discrepant excitation ratios.  If more than one
distinct \h2\ cloud exists in the sight line, their column densities can
sum and give anomalous column density ratios that are not reproducible by
one-sided slab models.  Large N(4)/N(2) can occur, for instance, if two
clouds are present: one cloud that contributes the majority of the \h2\
column density but no N(4), and one that contributes little N(\h2), little
N(2), but N(5) $\sim 10^{14.5}$ cm$^{-2}$.  This combination of clouds can
appear in Figure~\ref{ex_model} to have log[N(4)/N(2)]~=~$-1.0$, while
no single cloud in the model grid can achieve this high ratio. However,
it is unlikely that concatenation of multiple clouds will occur more
frequently in the LMC and SMC than in the Galaxy and thus explain their
anomalous column density ratios.

As we noted in the discussion of the \h2\ abundance, individual sight line
models are non-unique. However, taken together, the reduced abundance and
enhanced rotational excitation of the LMC and SMC \h2\ is strong evidence
for a shift in the formation/destruction balance in the low-metallicity
gas of the Clouds. Grain formation rates 1/3 to 1/10 times the Galactic
value and UV photo-absorption rates 10 to 100 times the Galactic mean
value are necessary to explain the observed patterns of \h2\ abundance
and rotational excitation.

\subsection{Dust and Gas Correlations} 

From the correlations of \h2\ and H~I in the samples, it is clear that
large increases in the fractional gas content of the Magellanic Clouds
relative to the Galaxy have not led to a corresponding increase in the
amount of molecular gas in the diffuse ISM. This fact, and the lower
dust-to-gas ratios in the Clouds, confirm that interstellar dust plays
a large role in the production of diffuse \h2. Thus, correlations with
the dust content of the clouds we probe may reveal such a link between
dust and \h2.

\begin{figure*}[t] 
\centerline{\epsfxsize=\hsize{\epsfbox{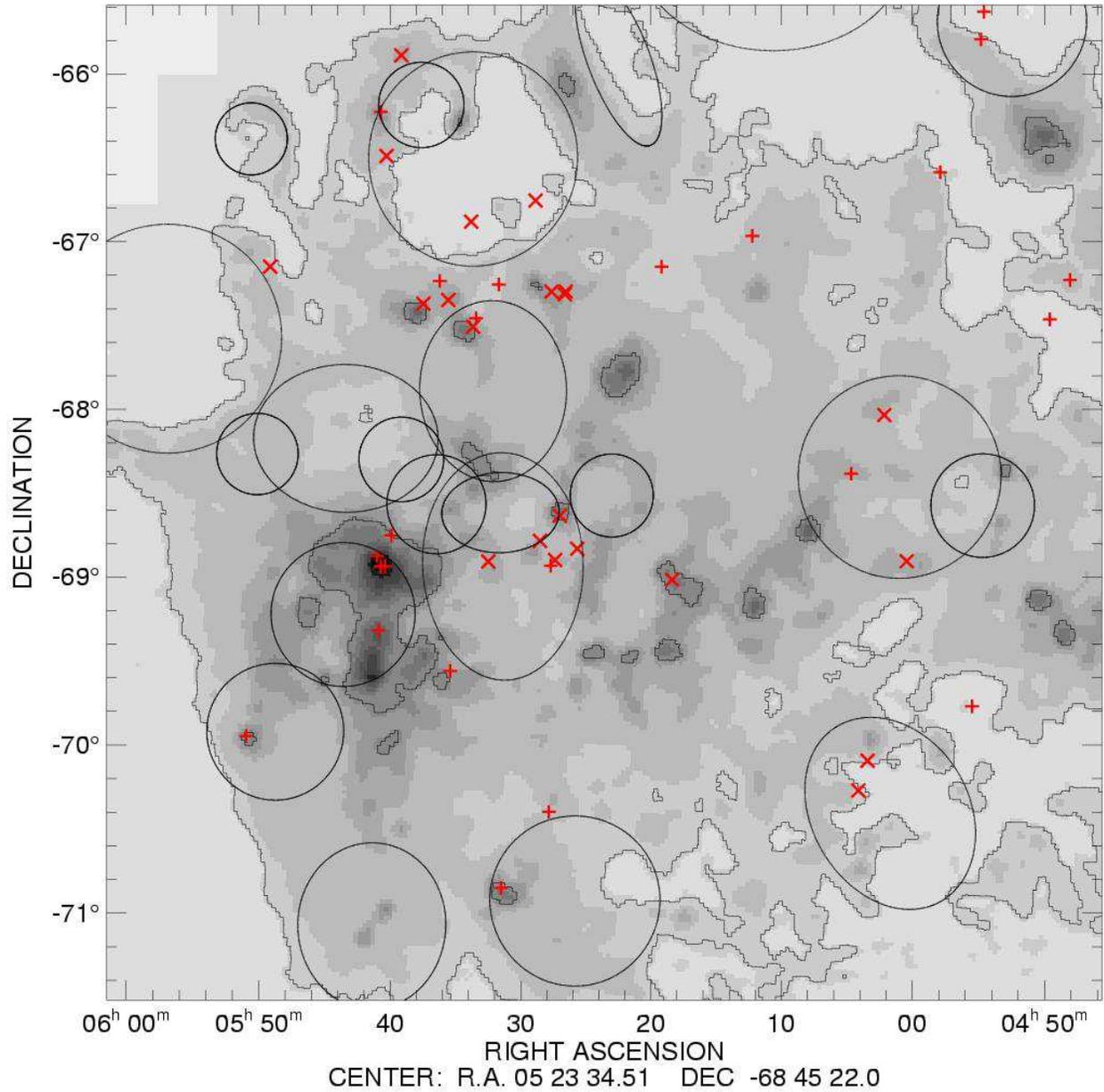}}}
\figcaption{\scriptsize IRAS 100 $\mu$m map of the LMC (Wheelock et al. 1991),
showing the survey targets.  A $+$ sign marks a detection, and $\times$
symbol designates a non-detection.  The contours are values of 10, 100,
and 1000 MJy sr$^{-1}$. In the LMC, there are 23 detections of \h2\ in 44
sight lines.  Aside from the strong detections of \h2\ around 30 Doradus 
(near 5$^{\rm h}$ 40$^{\rm m}$, -69$^{\circ}$),
there is no correlation between IRAS flux and \h2\ column density at a
given position. The ovals mark the positions of the 22 H~I supershells
identified by Kim et al.~(1999). The patchy nature of the LMC interstellar
medium and superbubble blowouts may explain the low \h2\ detection rate
in the LMC.
\label{lmcirasfig}} 
\end{figure*}
\begin{figure*}[t] 
\centerline{\epsfxsize=\hsize{\epsfbox{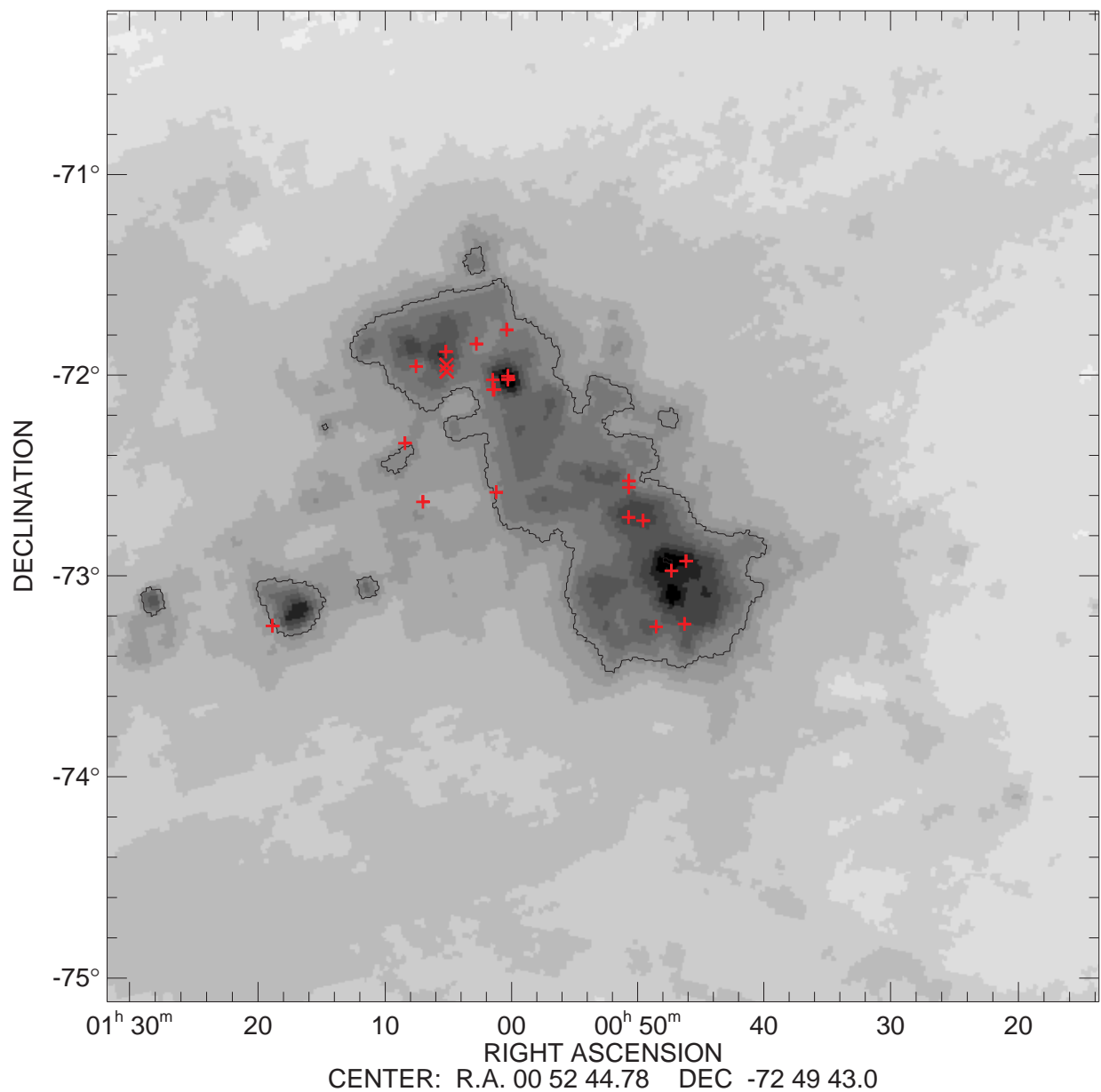}}}
\figcaption{\scriptsize IRAS 100 $\mu$m map of the SMC (Wheelock et al. 1991), showing
the locations of the survey targets.  A $+$ sign marks a detection of \h2,
and $\times$ symbol designates a non-detection.  The contours are values
of 10, 100, and 1000 MJy sr$^{-1}$. In the SMC, there are 24 detections
in 26 sight lines.
\label{smcirasfig}}
\end{figure*}

We use dust emission to test for a correlation between \h2\ and dust
abundance.  Figures~\ref{lmcirasfig} and \ref{smcirasfig} show IRAS 100
$\mu$m maps of the LMC and SMC, overlaid with surface brightness contours
at 10, 100, and 1000 MJy sr$^{-1}$.  Sight lines in which we detect \h2\
are marked with $+$, while sight lines for which we do not detect \h2\
are marked with $\times$.  For the LMC detections, the IRAS 100 $\mu$m
surface brightness ranges from 6 -- 2240 MJy sr$^{-1}$ with a mean of
277 MJy sr$^{-1}$. The LMC non-detections have 7 -- 146 MJy sr$^{-1}$
with a mean of 51 MJy sr$^{-1}$. For the SMC, the \h2\ detections have
IRAS brightnesses of 3 -- 110 MJy sr$^{-1}$ with a mean of 39 MJy sr$^{-1}$. The two
SMC non-detections have FIR surface brightness $\sigma_{\rm FIR}$ = 23
and 40 MJy sr$^{-1}$.  Thus, the observed far-infrared flux at a given
position is not a predictor of detectable \h2, except in regions of strong
dust emission such as 30 Doradus.  However, we note that the 1.5$'$ IRAS
pixel size subtends $\sim$ 22 pc at the distance of the LMC and 26 pc at
the SMC. Shull et al.~(2000) inferred linear sizes $d \simeq 2-3$ pc for
three Galactic \h2-bearing clouds. The different molecular excitation in
the HD 5980 and AV 232 sight lines, separated by 17 pc at the SMC, argues
for clouds sizes smaller than the size scale probed by the IRAS pixel.
These two techniques may sample different distributions of gas in the ISM,
and perhaps should not be expected to correlate.

\subsection{Diffuse \h2 Mass of the LMC and SMC}
 
Over the past decade, the Magellanic Clouds have been studied in
increasing detail in mm-wave CO emission.  In all cases, the results point
to CO emission weaker by factors of 3--5 than expected if Milky Way GMCs
were observed at a distance of 50--60 kpc.  Israel et al.\ (1982, 1986)
studied the LMC and SMC in $^{12}$CO (2-1) emission, using the ESO 3.6m
telescope at La Silla, Chile with a $2'$ FWHM beam at 230 GHz (1.3 mm).
They found relatively weak CO emission, with peak antenna temperatures
about 25\% of the expected signal from typical Galactic molecular clouds
at Magellanic Cloud distances.  They also found that the CO emission is
less widespread in the SMC, compared to either the Galaxy or the LMC.
More recent surveys with the ESO/SEST telescope at $45''$ resolution
(Israel et al.\ 1993; Rubio, Lequeux, \& Boulanger 1993) and the NANTEN
4 m telescope at 2.6$'$ resolution (Fukui et al.~1999) find similarly
weak $^{12}$CO (1-0) emission in the LMC and SMC.

The preliminary estimate for the calibration between $^{12}$CO line
intensity and total hydrogen column density in the SMC (Israel et
al.~1993) is approximately four times larger than the canonical
Galactic value, $X_{\rm gal}$ = [N(H~I)+2N(H$_2)]/I_{\rm CO} \approx
2 \times 10^{20}$ cm$^{-2}$ (K~km~s$^{-1})^{-1}$. For the LMC, Fukui
et al.~(1999) find a ratio $X_{\rm LMC} = 9 \times 10^{20}$ cm$^{-2}$
(K km s$^{-1}$)$^{-1}$, about half the value derived by Israel (1997)
using IRAS emission. The NANTEN survey estimates a total \h2\ cloud mass
of (4 - 7) $\times 10^{7}$ \Msun\ for clouds with N(\h2) $\geq 1 \times
10^{21}$ cm$^{-2}$, corresponding to 5 - 10\% of the atomic mass.

Most of these studies (Israel et al. 1986; Rubio et al.\ 1993) conclude
that the weak CO emission in the Magellanic Clouds does not arise from
a deficit of molecular gas, but rather from the low metallicity and high
UV radiation fields.  These factors result in low C and O abundances and
high CO photodissociation rates, which reduce the CO/H$2$ ratio from
the standard Galactic values. These suggestions have been sustained
by plane-parallel models of molecule formation and radiative transfer
(Maloney \& Black 1988; Lequeux et al.\ 1994).  In plane-parallel models
of clouds with N(H$_2) = 10^{22}$ cm$^{-2}$, Maloney \& Black (1988)
find that the GMCs with low (SMC) metallicity and correspondingly
reduced dust/gas ratios should have much smaller CO column densities
and CO-emitting surface areas. They also find that the H$_2$ number
densities are not as sensitive to the lower metallicities. Consequently,
the LMC and SMC molecular clouds are expected to show low CO/H$2$ ratios.
In this section, we use the FUSE data on H$_2$ to directly address the
issue of the H$_2$ mass fractions in the LMC and SMC.

Using the observed distribution of \h2\ and the high-resolution H~I
emission maps of Stanimirovic et al. (1999) and Staveley-Smith et al.
(2001), we can estimate the total mass of \h2 in the diffuse regions
of the LMC and SMC. To assign an \h2\ column density to each observed
position in the H~I column density map, we use a plot of N(\h2) versus
N(H~I) for the FUSE targets to produce 5 bins of $\Delta N_{HI}$ =
1 $\times 10^{21}$ cm$^{-2}$.  We assign a molecular fraction \fh2\
to each point in the bin, with \fh2\ randomly selected from the set of
FUSE targets with N(H~I) in that bin. This technique is used to produce
a synthetic \h2\ map. A direct integral over this map produces a total
diffuse \h2\ mass for the Clouds.

For the LMC, we obtain a total atomic hydrogen mass $4.8 \times 10^{8}$
\Msun.  From the H~I map and the observed FUSE distribution of \h2,
we obtain a molecular mass N(\h2) = $8 \times 10^{6}$ \Msun. Thus,
we find that $\sim$ 2\% of the hydrogen mass in the LMC resides in
diffuse \h2. This is consistent with the cumulative molecular fraction
for the LMC, $\langle f_{\rm H2} \rangle = 0.012$.  For the SMC, this
technique determines a total atomic hydrogen mass $4 \times 10^{8}$
\Msun.  We find a molecular mass N(\h2) = $2 \times 10^{6}$ \Msun.
Thus, 0.5\% of the hydrogen mass of the SMC lies in the diffuse \h2.

\begin{figure*}[t] 
\centerline{\epsfxsize=\hsize{\epsfbox{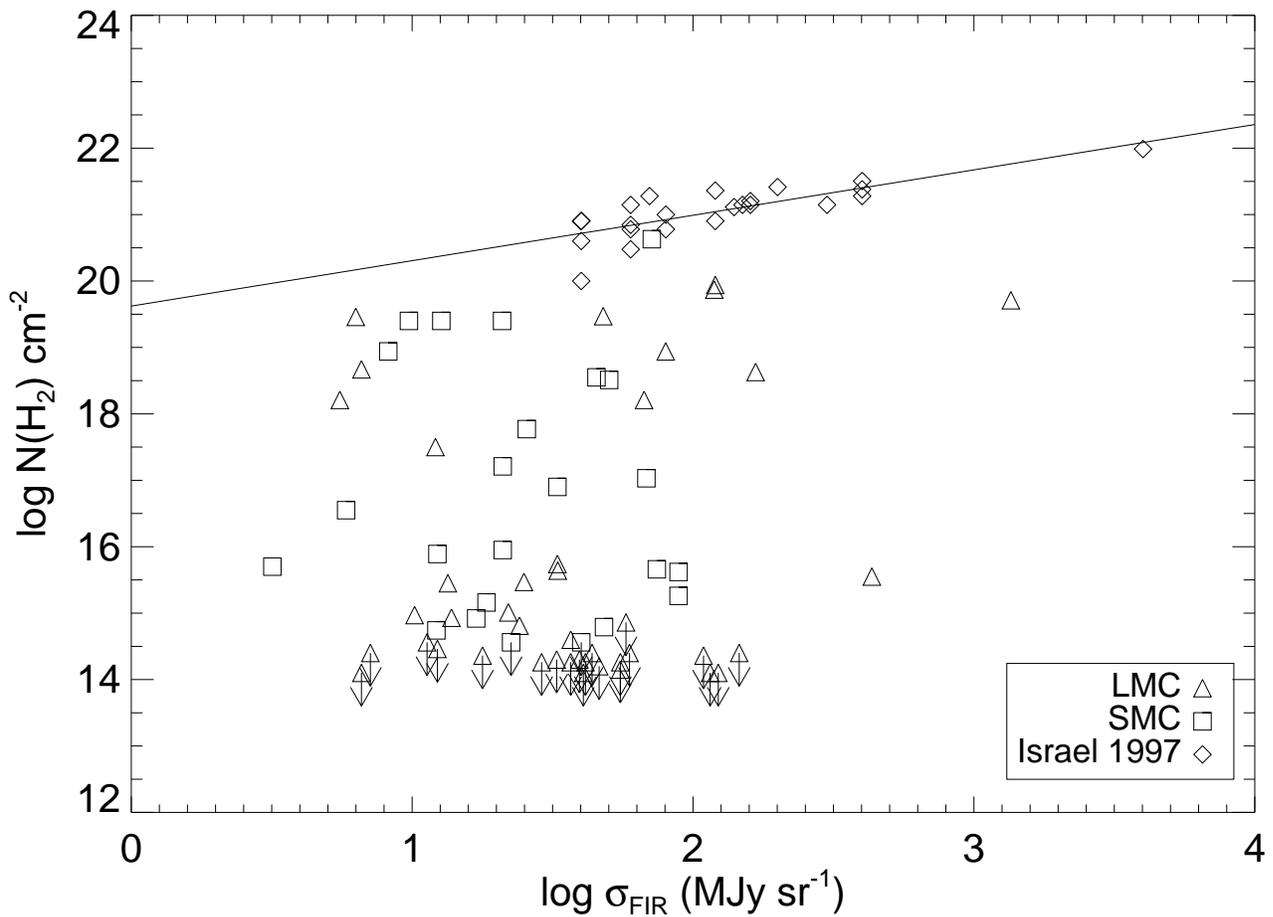}}}
\figcaption{\scriptsize No correlation is found between IRAS 100 $\mu$m surface
brightness, $\sigma_{\rm FIR}$, of the LMC CO clouds (Cohen et
al. 1988) and N(\h2) measured with FUSE.  The method of Israel (1997)
for deriving \h2\ column densities from far-infrared flux is compared to
the pencil-beam FUSE survey as discussed in the text. No correlation is
found between IRAS 100 $\mu$m flux and N(\h2) measured with FUSE. The
total \h2\ mass for the LMC extrapolated from the FUSE sample lies
$10\times$ below the mass inferred by Israel (1997). This indicates a
significantly lower molecular content in the LMC or a substantial cold,
dense component not sampled by the FUSE sample.
\label{corr_fir}}
\end{figure*}

Our inferred masses represent less than 1\% of the diffuse H~I measured
in the Clouds.  Israel (1997) used IRAS 100 $\mu$m maps and detected CO
clouds (Cohen et al.~1988) in the LMC to predict the surface density
of \h2\ in the LMC as a function of CO surface brightness in radio
measurements.  That study found total molecular masses of $(1.0 \pm
0.3) \times 10^{8}$ \Msun\ for the LMC and $(0.75 \pm 0.25) \times
10^{8}$ \Msun\ for the SMC.  These masses correspond to $\sim 20$\%
mass fraction in the Clouds, 10 times that found by the FUSE survey.
As noted above, the NANTEN survey (Fukui et al.~1999) found values about
half this amount for the LMC.  We compare the detected column densities
of \h2\ to these predictions in Figure~\ref{corr_fir}.  Although we do
not probe directly the CO clouds on which the Israel (1997)  and Fukui
et al.~(1999) results are indirectly based, the FUSE target stars do
coincide with regions of similar IRAS 100 $\mu$m surface brightness. We
find that total H content in our FUSE sight lines is well correlated with
the FIR surface brightness, but in the range of overlap, our detected \h2\
column densities lie 1-5 orders of magnitude below that predicted by the
IRAS flux. Only one FUSE sight line coincides with the predicted surface
density, indicating that the total molecular mass in the LMC may be far
lower than that predicted by the total dust and/or CO content. Based on
Figure~\ref{corr_fir} we conclude that the relationships between N(CO),
N(H~I), and FIR surface brightness that predict high N(\h2) do not apply
in diffuse regions away from the detected CO clouds. These regions,
which we probe directly with the FUSE sight lines, possess a wide range
of molecular fraction but are generally below the value \fh2 $\sim 0.1$
required to produce M(\h2) / M(H~I) = 0.2.

We derived M(\h2) for the Clouds by assigning observed molecular
fractions to the 21 cm emission maps according to the observed H~I
column density. In both the SMC and LMC maps, a small fraction of the
beam positions have measured N(H~I) larger than any H~I column probed in
this survey. In the LMC, the maximum N(H~I) exceeds the highest value in
our survey by 2$\times$, and positions with N(H~I) $> 5.0 \times 10^{21}$
cm$^{-2}$ comprise 40\% of the H~I by mass. In the SMC, N(H~I) exceeds the
survey maximum by 30\%, but these positions comprise only 0.4\% of the
H~I by mass. Thus, in the SMC, this survey has sampled the distribution
of observed N(H~I) rather completely, while in the LMC the upper end
of the distribution has not been sampled. 

The discrepancy between \h2\ masses determined from FUSE and those
predicted from the 100 $\mu$m flux can be resolved if the H~I positions
with N(H~I) higher than our sample are highly molecular.  However, the
molecular fractions in these points must be extremely high to achieve
M(\h2) $\sim 10^{8}$ \Msun. For the SMC, those points with N(H~I) $>
8.0 \times 10^{21}$ cm$^{-2}$ must have \fh2\ = 0.87 to give M(\h2) =
$7.5 \times 10^{7}$ \Msun. For the LMC, where N(H~I) ranges to
2$\times$ that in our sample, \fh2\ = 0.70 is required. These molecular
fractions are substantially higher than the \fh2\ $\sim$ 0.1 seen in
high column-density Galactic molecular clouds and are somewhat like the
\fh2\ $\sim$ 0.7 seen in Galactic translucent clouds (Snow et al.~2000;
Rachford et al.~2001) or GMCs. Thus the molecular mass discrepancies
are resolved since our survey does not sample the upper end of the
molecular cloud regime.  A large fraction of \h2\ could reside in cold,
dense gas that appears in neither the H~I emission map nor in the FUSE
data, which probes to E$'$(B-V) = 0.20 only. This question will be
addressed by FUSE Cycle 2 observations of seven sight lines to the SMC
and LMC with E(B-V) $\geq 0.20$.

\section{Discussion and Conclusions} 

We have surveyed the abundance, excitation, and properties of diffuse \h2\
in the diffuse interstellar medium of the Small and Large Magellanic
Clouds. The central results of this survey are the column density
measurements themselves (Tables 5 and 6) and the statistics of \h2\
detections and non-detections. In addition to these basic findings, we
have also correlated the detected \h2\ with the H~I and dust along these
sight lines, and derived rotational excitation parameters of the \h2.

Our major conclusions are: 
\begin{itemize} 
\item In a sample of 70 sight lines to hot stars in the LMC and SMC, we find 
      \h2\ towards 23 of 44 LMC stars and towards 24 of 26 SMC stars. 

\item The overall abundance of diffuse \h2\ in the ISM of the
      Magellanic Clouds is lower than in the Milky Way.  We find
      molecular fractions $\langle f_{\rm H2} \rangle =
      0.010^{+0.005}_{-0.002}$ for the SMC and $\langle f_{\rm H2}
      \rangle = 0.012^{+0.006}_{-0.003}$ for the LMC, compared with
      $\langle f_{\rm H2} \rangle = 0.095$ for the Galactic disk over a
      similar range of reddening. The main source of error is
      systematic uncertainty in the measurement of N(H~I). This result
      can be interpreted as the effect of suppressed \h2\ formation on
      dust grains and enhanced \h2\ destruction by FUV photons.

\item The reduced molecular fraction and enhanced rotational excitation
      in the Clouds are consistent with models having a low \h2\
      formation rate coefficient and a high UV photo-absorption rate,
      relative to Galactic conditions.  Using grids of models clouds,
      we find that $R \sim 3 \times 10^{-18}$ cm$^{3}$ s$^{-1}$,
      1/3 - 1/10 the Galactic rate, and $I \sim 10^{-7} - 10^{-6}$ photons
      cm$^{-2}$ s$^{-1}$ Hz$^{-1}$, 10 to 100 times the Galactic mean,
      are necessary to replicate the LMC and SMC results.

\item The mean molecular fraction of the detected diffuse gas implies a
      total diffuse \h2\ mass of M(\h2) = $8 \times 10^{6}$
      \Msun\ for the LMC and M(\h2) $= 2 \times 10^{6}$ \Msun\
      for the SMC. These masses are $\lesssim 2$\% of the H~I masses of
      the Clouds, implying small overall molecular content, high star
      formation efficiency, and/or substantial molecular mass in cold,
      dense clouds unseen by our survey.

\item We find no correlation between observed \h2\ and dust properties
      below a critical value of E(B-V) and/or a critical 100 $\mu$m
      IRAS flux. However, high extinction and/or strong dust emission
      are predictors of high \h2\ abundance.

\end{itemize} 
 

This survey is one of four major surveys of interstellar \h2\ that are
being conducted by the Colorado group with FUSE.  The Galactic \h2\
survey consists of over 100 sight lines to hot stars in the disk, with
$|b| < 10^{\circ}$.  The Galactic halo survey consists of $\sim$100
sight lines to the Magellanic Clouds and extragalactic AGN and QSOs.
Finally, the FUSE translucent cloud survey consists of 35 stars in the
Galactic disk, with $A_V \gtrsim 2.0$. Taken together, these surveys
and the detailed followups should form the basis of a more complete
understanding of molecule formation in the diffuse ISM than has been
possible to date.

\acknowledgements

Lister Staveley-Smith graciously provided the LMC H~I maps in advance
of publication.  Comments from Bill Blair, Phil Maloney, Philipp Richter,
and Dan Welty improved the manuscript.  We acknowledge the use of NASA's
{\em SkyView} facility (http://skyview.gsfc.nasa.gov) located at NASA
Goddard Space Flight Center.  We have also made extensive use of the
SIMBAD database, operated at CDS, Strasbourg, France.  This work was
supported in part by the Colorado astrophysical theory program through
NASA grants NAG5-4063 and NAG5-7262.

\begin{deluxetable}{llllclccc} 
\tablecolumns{9} 
\tablenum{1} 
\tablewidth{0pt} 
\tablecaption{FUSE \h2\ Large Magellanic Cloud Target List} 
\tablehead{ 
\colhead{Name} 
&\colhead{RA (J2000)} 
&\colhead{Dec (J2000)} 
&\colhead{Spectral Type} 
&\colhead{Ref.\tablenotemark{a}} 
&\colhead{Observation}  
&\colhead{Exposures} 
&\colhead{$t_{\rm exp}$} 
&\colhead{S/N} \\  
\colhead{} 
&\colhead{h m s} 
&\colhead{d m s} 
&\colhead{} 
&\colhead{} 
&\colhead{ID} 
&\colhead{(used/total)} 
&\colhead{(ks)} 
&\colhead{(per pixel)\tablenotemark{b}}  
} 
 
\startdata 
Sk -67 05       & 4 50 18.96 & -67 39 37.90     & O9.7 Ib           &  1 & P1030703   & 1/1  & 8.0  & 10 \\
Sk -67 14       & 4 54 31.92 & -67 15 24.90     & B1.5 Ia           &  4 & P1174201,3 & 3/5  & 10.5 & 7  \\
HD 32109        & 4 55 31.50 & -67 30 1.00      & WN4b              &  5 & P1174401-2 & 5/5  & 16.7 & 8  \\
Sk -65 21       & 5 1 22.33  & -65 41 48.10     & O9.7 Iab          &  8 & P1030901-4 & 4/4  & 18.2 & 14 \\
Sk -65 22       & 5 1 23.14  & -65 52 33.50     & O6 Iaf$+$         &  1 & P1031002   & 11/11& 27.2 & 9  \\
HD 33133        & 5 3 9.00   & -66 40 57.76     & WN8h              &  5 & P1174501   & 4/4  & 4.6  & 7  \\
Sk -69 59       & 5 3 12.79  & -69 1 37.20      & B0                &  2 & P1031103   & 4/4  & 25.7 & 9  \\
Sk -70 60       & 5 4 40.94  & -70 15 34.50     & O4-5 V:n          &  9 & P1172001   & 2/2  & 7.9  & 6  \\
Sk -70 69       & 5 5 18.73  & -70 25 49.80     & O5 V              &  8 & P1172101   & 2/2  & 6.1  & 6  \\
Sk -68 41       & 5 5 27.20  & -68 10 2.70      & B0.5 Ia           &  4 & P1174102   & 1/1  & 6.5  & 6  \\
Sk -68 52       & 5 7 20.60  & -68 32 9.60      & B0 Ia             &  1 & P1174001   & 2/5  & 8.1  & 7  \\
Sk -67 69       & 5 14 20.16 & -67 8 3.50       & O4 III(f)         &  7 & P1171703   & 4/4  & 7.8  & 7  \\
Sk -69 104      & 5 18 59.57 & -69 12 54.70     & O6 Ib(f)          &  1 & P1172401   & 1/1  & 4.1  & 9  \\
Sk -67 76       & 5 20 5.81  & -67 21 8.90      & B0                &  2 & P1031201   & 3/5  & 19.2 & 9  \\
Sk -69 124      & 5 25 18.37 & -69 3 11.10      & O9 Ib             & 10 & P1173601-2 & 3/3  & 12.7 & 9  \\
Sk -67 101      & 5 25 56.36 & -67 30 28.70     & O8 II((f))        &  9 & P1173402   & 2/2  & 6.2  & 6  \\
Sk -67 104      & 5 26 4.15  & -67 29 56.50     & WC4(+O?)+O8I:     &  3 & P1031302   & 3/3  & 5.1  & 9  \\
Sk -68 80       & 5 26 30.43 & -68 50 26.60     & WC4+O6V-III       &  3 & P1031402   & 4/4  & 9.7  & 9  \\
BI 170          & 5 26 47.79 & -69 6 11.70      & O9.5 Ib           &  9 & P1173701   & 1/1  & 4.3  & 7  \\
Sk -67 111      & 5 26 48.00 & -67 29 33.00     & O6: Iafpe         &  9 & P1173001   & 2/2  & 8.0  & 8  \\
BI 173          & 5 27 10.08 & -69 7 56.20      & O8 II:            &  9 & P1173201-2 & 4/4  & 11.4 & 8  \\
Sk -70 91       & 5 27 33.74 & -70 36 48.30     & O6.5 V            & 10 & P1172501   & 1/1  & 5.5  & 7  \\
Sk -66 100      & 5 27 45.59 & -66 55 15.00     & O6 II(f)          &  8 & P1172303   & 2/2  & 7.1  & 7  \\
HD 269582       & 5 27 52.75 & -68 59 8.60      & WN10h             & 11 & P1174701   & 1/1  & 4.6  & 7  \\
HD 37026        & 5 30 12.22 & -67 26 8.40      & WC4               & 12 & P1175101   & 3/3  & 8.1  & 6  \\
Sk -71 45       & 5 31 15.55 & -71 4 8.90       & O4-5 III(f)       &  1 & P1031501-4 & 4/4  & 18.9 & 15 \\
HD 269687       & 5 31 25.61 & -69 5 38.40      & WN11h             & 11 & P1174801   & 1/1  & 3.8  & 6  \\
Sk -67 166      & 5 31 42.00 & -67 38 6.00      & O4 If+            &  1 & A1330100   &  84  & 220. & 65 \\ 
Sk -67 169      & 5 31 51.68 & -67 2 22.30      & B1 Ia             &  4 & P1031603   & 4/7  & 40.5 & 8  \\
Sk -67 167      & 5 31 51.98 & -67 39 41.10     & O4 Inf+           &  7 & P1171902   & 1/1  & 2.8  & 10 \\
Sk -67 191      & 5 33 34.12 & -67 30 19.60     & O8 V              & 10 & P1173102   & 1/1  & 6.6  & 6  \\
BI 208          & 5 33 57.45 & -67 24 20.00     & O7 Vz             &  9 & P1172702-5 & 4/4  & 15.6 & 9  \\
HD 37680        & 5 34 19.39 & -69 45 10.00     & WC4               & 12 & P1175001   & 3/3  & 7.0  & 5  \\
HD 269810       & 5 35 13.92 & -67 33 27.00     & O2 III(f\,$^*$)   &  6 & P1171603   & 5/5  & 8.2  & 9  \\
BI 229          & 5 35 32.20 & -66 2 37.60      & O7 V-III          &  9 & P1172801   & 2/2  & 5.6  & 7  \\
Sk -66 169      & 5 36 54.50 & -66 38 25.00     & O9.7 Ia+          &  4 & P1173801   & 3/3  & 4.8  & 8  \\
Sk -66 172      & 5 37 5.56  & -66 21 35.70     & O2 III(f\,$^*$)+OB&  6 & P1172201   & 1/1  & 3.6  & 6  \\
Sk -68 135      & 5 37 48.60 & -68 55 8.00      & ON9.7 Ia+         &  1 & P1173901   & 3/3  & 6.8  & 5  \\
Mk 42           & 5 38 42.10 & -69 5 54.70      & O3 If\,$^*$/WN6-A & 14 & P1171802-4 & 5/5  & 15.4 & 15 \\
Sk -69 243      & 5 38 42.57 & -69 6 3.20       & WN+OB composite   & 13 & P1031705-6 & 2/2  & 9.5  & 12 \\
Sk -69 246      & 5 38 53.50 & -69 2 0.70       & WN6h              &  5 & P1031802   & 4/4  & 22.1 & 9  \\
HD 269927       & 5 38 58.25 & -69 29 19.10     & WN9h              &  5 & P1174601   & 2/2  & 7.2  & 8  \\
BI 272          & 5 44 23.18 & -67 14 29.30     & O7: III-II:       &  9 & P1172902   & 1/1  & 6.6  & 7  \\
Sk -70 115      & 5 48 49.76 & -70 3 57.50      & O6.5 Iaf          &  9 & P1172601   & 2/2  & 4.8  & 7  \\
\enddata 
\tablenotetext{a}{Spectral type references: (1) Walborn (1977); (2)
Rousseau et al. (1978); (3) Moffat et al.~(1990); (4) Fitzpatrick (1988);
(5) Smith et al.~(1996); (6) Walborn et al.~(2001b); (7) Garmany \& Walborn
(1987); (8) Walborn et al.~(1995); (9) Walborn et al.~(2001a) (10) Conti
et al. (1986); (11) Crowther \& Smith (1997); (12) Torres-Dodgen \& Massey
(1988); (13) Breysacher et al.~(1999); (14) Walborn \& Blades~1997.} 
\tablenotetext{b}{Signal-to-noise ratio per pixel in the 1040-1050 
\AA\ range. Signal-to-noise ratio per resolution element varies 
with the spectral resolution, which is not fixed in our survey.}  
\end{deluxetable} 

\begin{deluxetable}{llllclccc} 
\tablecolumns{9} 
\tablenum{2} 
\tablewidth{0pt} 
\tablecaption{FUSE \h2\ Small Magellanic Cloud Target List} 
\tablehead{ 
\colhead{Name} 
&\colhead{RA (J2000)} 
&\colhead{Dec (J2000)} 
&\colhead{Spectral Type} 
&\colhead{Ref.\tablenotemark{a}} 
&\colhead{Observation}  
&\colhead{Exposures} 
&\colhead{$t_{\rm exp}$} 
&\colhead{S/N} \\  
\colhead{} 
&\colhead{h m s} 
&\colhead{d m s} 
&\colhead{} 
&\colhead{} 
&\colhead{ID} 
&\colhead{(used/total)} 
&\colhead{(ks)} 
&\colhead{(per pixel)\tablenotemark{b}}  
} 
\startdata 
AV 14           & 0 46 32.66 & -73 6 5.60       & O3-4 V           &  1 & P1175301  & 2/2  &  6.8 & 5  \\ 
AV 15           & 0 46 42.19 & -73 24 54.70     & O6.5 II(f)       &  2 & P1150101  & 12/12& 14.6 & 7  \\ 
AV 26           & 0 47 50.07 & -73 8 20.70      & O7 III           &  1 & P1176001  & 1/1  &  4.0 & 4  \\ 
AV 47           & 0 48 51.35 & -73 25 57.60     & O8 III((f))      &  2 & P1150202  & 5/5  & 16.3 & 8  \\ 
AV 69           & 0 50 17.40 & -72 53 29.90     & OC7.5 III((f))   &  2 & P1150303  & 13/13& 17.6 & 9  \\ 
AV 75           & 0 50 32.50 & -72 52 36.20     & O5 III(f$+$)     &  2 & P1150404  & 4/4  & 14.4 & 7  \\  
AV 83           & 0 50 52.01 & -72 42 14.50     & O7 Iaf$+$        &  2 & P1176201  & 1/1  &  4.0 & 5  \\ 
AV 95           & 0 51 21.54 & -72 44 12.90     & O7 III((f))      &  2 & P1150505  & 11/11& 14.3 & 7  \\ 
AV 207          & 0 58 33.19 & -71 55 46.50     & O7 V             &  3 & P1175901  & 3/3  &  4.1 & 3  \\ 
NGC 346-6       & 0 58 57.74 & -72 10 33.60     & O4 V((f))        &  6 & P1175601  & 4/4  &  6.4 & 8  \\ 
NGC 346-4       & 0 59 0.39  & -72 10 37.90     & O5-6 V           &  5 & P1175701  & 2/2  &  5.2 & 8  \\ 
NGC 346-3       & 0 59 1.09  & -72 10 28.20     & O2 III(f\,$^*$)  & 12 & P1175201  & 2/2  &  5.1 & 8  \\ 
HD 5980         & 0 59 26.57 & -72 9 53.90      & WN var           &  4 & X0240202  & 8/8  &  3.2 & 9  \\ 
AV 232          & 0 59 32.19 & -72 10 46.20     & O7 Iaf$+$        &  8 & X0200201  & 5/5  & 10.3 & 8  \\ 
Sk 82           & 0 59 45.72 & -72 44 56.10     & B0 Iaw           &  9 & P1030301  &10/10 & 16.2 & 8  \\ 
AV 238          & 0 59 55.61 & -72 13 37.70     & O9.5 III         &  2 & P1176601  & 2/2  & 11.1 & 6  \\ 
AV 242          & 1 0 6.84   & -72 13 57.00     & B1 Ia            &  7 & P1176901  & 2/2  &  5.1 & 6  \\ 
AV 264          & 1 1 7.72   & -71 59 58.60     & B1 Ia            &  7 & P1177001  & 3/3  &  4.5 & 5  \\ 
AV 321          & 1 2 57.04  & -72 8 9.30       & B0 IIIww         &  1 & P1150606  & 6/6  & 16.9 & 7  \\ 
AV 327          & 1 3 10.58  & -72 2 13.80      & O9.5 II-Ibw      &  2 & P1176401  & 4/4  &  4.7 & 6  \\ 
Sk 108          & 1 3 25.24  & -72 6 43.30      & WN3 + O6.5(n)    &  8 & X0150102  & 8/8  & 13.3 & 4  \\ 
AV 372          & 1 4 55.73  & -72 46 47.70     & O9.5 Iabw        & 10 & P1176501  & 1/1  &  4.4 & 5  \\ 
AV 378          & 1 5 9.44   & -72 5 35.00      & O8 V             &  1 & P1150707  & 8/8  & 14.7 & 7  \\ 
AV 388          & 1 5 39.62  & -72 29 26.80     & O4 V             &  1 & P1175401  & 2/2  &  5.6 & 6  \\ 
Sk 159          & 1 15 58.84 & -73 21 24.10     & B0.5 Iaw         &  9 & P1030501  & 2/2  &  6.9 & 7  \\ 
Sk 188          & 1 31 4.23  & -73 25 2.20      & WO3+O4V          & 11 & P1030601-4& 8/8  & 23.1 & 15 \\ 
\enddata 
\tablenotetext{a}{Spectral type references: (1)  Garmany et al.~(1987);
(2)  Walborn et al.~(2000); (3)  Crampton \& Greasley (1982); (4)
Koenigsberger et al.~(1994); (5) Walborn \& Blades (1986); (6) Walborn et
al.~(1995); (7) Lennon (1997); (8)  Walborn (1977); (9)  Walborn (1983);
(10) Walborn et al.~(2001a); (11) Crowther et al.~(1998);
(12) Walborn et al.~(2001b).}
\tablenotetext{b}{Signal-to-noise ratio per pixel in 1040-1050 
\AA\ range. Signal-to-noise ratio per resolution element varies  
with spectral resolution, which is not fixed in our survey.}  
\end{deluxetable} 

\begin{deluxetable}{lcccccc}
\tablecolumns{7}
\tablenum{3}
\tablewidth{0pt}
\tablecaption{FUSE \h2\ Large Magellanic Cloud Summary} 
\tablehead{
\colhead{Name}
&\colhead{E(B-V)\tablenotemark{a}} 
&\colhead{$v_{\rm LSR}$\tablenotemark{b}}
&\colhead{N(H$_2$)}
&\colhead{$T_{01}$}
&\colhead{$b_{\rm Dopp}$}
&\colhead{N(H I)\tablenotemark{c}} \\ 
\colhead{}
&\colhead{}
&\colhead{(km s$^{-1}$)} 
&\colhead{(cm$^{-2}$)} 
&\colhead{(K)} 
&\colhead{(km s$^{-1}$)} 
&\colhead{($10^{20}$ cm$^{-2}$)} 
} 
\startdata
Sk -67 05  & 0.16 & 256 $\pm$ 7 & 19.46$^{+0.05}_{-0.05}$ & 57  & 6.2$^{+1.4}_{-1.1}$   & 7.7    \\
Sk -67 14  & 0.08 & 268 $\pm$ 3 & 15.01$^{+0.24}_{-0.18}$ & 270 & linear                & 8.7    \\
HD 32109   & 0.05 & 284 $\pm$ 13& 18.67$^{+0.22}_{-0.26}$ & 72  & 5.8$^{+1.8}_{-3.6}$   & 5.6    \\
Sk -65 21  & 0.12 & 222 $\pm$ 11& 18.21$^{+0.14}_{-0.35}$ & 90  & 2.3$^{+2.1}_{-1.7}$   & 5.4    \\
Sk -65 22  & 0.15 & 269 $\pm$ 4 & 14.93$^{+0.19}_{-0.16}$ & 117 & linear                & 10.3   \\
HD 33133   & 0.08 & 299 $\pm$ 14& 17.50$^{+0.26}_{-1.04}$ & 30  & 3.5$^{+2.6}_{-1.4}$   & 8.6    \\
Sk -69 59  & 0.12 & \nd         & $\leq$14.10             & \nd & \nd                   & 10.2   \\
Sk -70 60  & 0.135& \nd         & $\leq$14.46             & \nd & \nd                   & 3.5    \\
Sk -70 69  & 0.09 & \nd         & $\leq$14.56             & \nd & \nd                   & 3.3    \\
Sk -68 41  & 0.105& \nd         & $\leq$14.30             & \nd & \nd                   & 10.8   \\
Sk -68 52  & 0.17 & 232 $\pm$ 12& 19.47$^{+0.06}_{-0.05}$ & 60  & 4.8$^{+2.9}_{-3.6}$   & 12.8   \\
Sk -67 69  & 0.15 & 278 $\pm$ 11& 15.47$^{+1.51}_{-0.12}$ & 115 & 6.9$^{+3.3}_{-4.0}$   & 12.5   \\
Sk -67 104 & 0.13 & \nd         & $\leq$14.15             & \nd & \nd                   & 13.6   \\
Sk -67 76  & 0.11 & 247 $\pm$ 8 & 14.97$^{+0.40}_{-0.33}$ & 120 & linear                & 7.9    \\
Sk -69 124 & 0.125& \nd         & $\leq$14.26             & \nd & \nd                   & 8.3    \\
Sk -67 101 & 0.115& \nd         & $\leq$14.26             & \nd & \nd                   & 13.4   \\
Sk -69 104 & 0.105& \nd         & $\leq$14.40             & \nd & \nd                   & 13.8   \\
Sk -68 80  & 0.10 & \nd         & $\leq$14.10             & \nd & \nd                   & 7.1    \\
BI 170     & 0.125& \nd         & $\leq$14.26             & \nd & \nd                   & 7.2    \\
Sk -67 111 & 0.115& \nd         & $\leq$14.86             & \nd & \nd                   & 13.9   \\
BI 173     & 0.16 & 217 $\pm$ 11& 15.64$^{+0.43}_{-0.15}$ & 117 & 5.9$^{+2.2}_{-1.7}$   & 6.2    \\
Sk -70 91  & 0.10 & 281 $\pm$ 8 & 14.81$^{+0.63}_{-0.35}$ & 156 & linear                & 6.2    \\
Sk -66 100 & 0.11 & \nd         & $\leq$14.40             & \nd & \nd                   & 3.2    \\
HD 269582  & 0.09 & \nd         & $\leq$14.30             & \nd & \nd                   & 6.8    \\
HD 37026   & 0.11 & 246 $\pm$ 9 & 15.45$^{+1.94}_{-0.25}$ & 91  & 5.4$^{+3.0}_{-4.0}$   & 7.3    \\
Sk -71 45  & 0.18 & 228 $\pm$ 9 & 18.63$^{+0.09}_{-0.19}$ & 98  & 1.6$^{+2.8}_{-0.5}$   & 15.0   \\
HD 269687  & 0.10 & \nd         & $\leq$14.40             & \nd & \nd                   & 11.2   \\
Sk -67 166 & 0.10 & 209 $\pm$ 2 & 15.74$^{+1.16}_{-0.34}$ & 55  & 2.7$^{+0.7}_{-1.3}$   & 12.9   \\
Sk -67 169 & 0.07 & \nd         & $\leq$14.10             & \nd & \nd                   & 2.3    \\
Sk -67 167 & 0.12 & \nd         & $\leq$14.36             & \nd & \nd                   & 14.0   \\
Sk -67 191 & 0.11 & \nd         & $\leq$14.40             & \nd & \nd                   & 12.6   \\
BI 208     & 0.03 & 265 $\pm$ 11& 14.60$^{+0.27}_{-0.23}$ & 425 & linear                & 10.1   \\
HD 37680   & 0.15 & 222 $\pm$ 11& 18.94$^{+0.06}_{-0.19}$ & 65  & 3.7$^{+2.9}_{-1.4}$   & 15.9   \\
HD 269810  & 0.10 & \nd         & $\leq$14.10             & \nd & \nd                   & 15.3   \\
BI 229     & 0.135 & \nd        & $\leq$14.26             & \nd & \nd                   & 9.3    \\
Sk -66 169 & 0.14 & \nd         & $\leq$14.20             & \nd & \nd                   & 6.1    \\
Sk -66 172 & 0.19 & 267 $\pm$ 9 & 18.21$^{+0.39}_{-0.32}$ & 41 & 6.0$^{+2.4}_{-2.1}$    & 10.3   \\
Sk -68 135 & 0.27 & 250 $\pm$ 5 & 19.87$^{+0.07}_{-0.07}$ & 91 & 4.3$^{+2.4}_{-2.1}$    & 27.8   \\
Sk -69 246 & 0.10 & 261 $\pm$ 9 & 19.71$^{+0.03}_{-0.03}$ & 74  & 4.8$^{+4.0}_{-0.5}$   & 35.2   \\
HD 269927  & 0.19 & 257 $\pm$ 7 & 15.55$^{+0.27}_{-0.11}$ & 163 & 8.1$^{+3.9}_{-3.2}$   & 31.9   \\
BI 272     & 0.14 & \nd         & $\leq$14.36             & \nd & \nd                   & 6.3    \\
Sk -70 115 & 0.21 & 202 $\pm$ 4 & 19.94$^{+0.07}_{-0.07}$ & 53  & 2.3$^{+1.4}_{-0.8}$   & 23.9   \\
\enddata
\tablenotetext{a}{Color excess is total of Galactic plus LMC.
Corrections are applied to correlations in the text. } 
\tablenotetext{b}{LSR velocity assuming that the Galactic \h2\ lines lie at 
v$_{LSR}$ = 0 km s$^{-1}$.} 
\tablenotetext{c}{The tabulated N(H~I) are corrected for
the systematic effects discussed in \S~2.4. The values here are N(H~I) = (0.75 $\pm$ 0.25) N(21 cm).}
\end{deluxetable}
\normalsize

\begin{deluxetable}{lcccccc}
\tablecolumns{7}
\tablenum{4}
\tablewidth{0pt}
\tablecaption{FUSE \h2\ Small Magellanic Cloud Summary}
\tablehead{
\colhead{Name}
&\colhead{E(B-V)\tablenotemark{a}} 
&\colhead{$v_{\rm LSR}$\tablenotemark{b}}
&\colhead{N(H$_2$)}
&\colhead{$T_{01}$}
&\colhead{$b_{\rm Dopp}$}
&\colhead{N(H I)\tablenotemark{c}} \\ 
\colhead{}
&\colhead{}
&\colhead{(km s$^{-1}$)} 
&\colhead{(cm$^{-2}$)} 
&\colhead{(K)} 
&\colhead{(km s$^{-1}$)} 
&\colhead{($10^{20}$ cm$^{-2}$)} 
} 
\startdata
AV 14    & 0.135& $124 \pm 11$ & 17.03$^{+1.35}_{-0.76}$ & 53  &      14.3$^{+5.6}_{-5.7}$     & 80.3   \\
AV 15    & 0.10 & $130 \pm 7$  & 16.90$^{+1.14}_{-0.89}$ & 47  &      9.1$^{+3.0}_{-3.0}$      & 58.2   \\
AV 26    & 0.13 & $122 \pm 9$  & 20.63$^{+0.05}_{-0.05}$ & 53  &      1.8$^{+5.9}_{-1.3}$      & 81.8   \\
AV 47    & 0.075& $117 \pm 8$  & 17.77$^{+0.34}_{-1.13}$ & 363 &      6.1$^{+3.0}_{-2.0}$      & 48.0   \\
AV 69    & 0.11 & $112 \pm 8$  & 18.73$^{+0.16}_{-0.15}$ & 90  &      7.2$^{+2.1}_{-2.2}$      & 80.3   \\
AV 75    & 0.155& $108 \pm 11$ & 18.51$^{+0.21}_{-0.62}$ & 47  &      9.6$^{+3.0}_{-2.1}$      & 80.3   \\
AV 83    & 0.12 & $107 \pm 7$  & 15.16$^{+0.11}_{-0.10}$ & 125 &      14.5$^{+5.0}_{-8.0}$     & 27.5   \\
AV 95    & 0.06 & $107 \pm 7$  & 19.40$^{+0.08}_{-0.07}$ & 78  &      10.3$^{+3.9}_{-2.0}$     & 44.6   \\
AV 207   & 0.105& $152 \pm 10$ & 19.40$^{+0.08}_{-0.09}$ & 78  &      2.7$^{+6.1}_{-1.2}$      & 32.3   \\
NGC346-3 & 0.095& $130 \pm 9$  & 15.62$^{+1.38}_{-0.10}$ & 61  &      17.8$^{+\infty}_{-3.0}$  & 78.0   \\
NGC346-4 & 0.095& $129 \pm 11$ & 15.37$^{+0.57}_{-0.06}$ & 90  &      19.0$^{+\infty}_{-3.0}$  & 78.0   \\
NGC346-6 & 0.095& $131 \pm 6$  & 15.74$^{+0.42}_{-0.10}$ & 105 &      17.3$^{+5.9}_{-7.4}$     & 78.0   \\
HD 5980  & 0.07 & $124 \pm 10$ & 15.66$^{+0.26}_{-0.13}$ & 101 &      9.2$^{+3.0}_{-2.0}$      & 26.1   \\
AV 232   & 0.11 & $119 \pm 10$ & 15.26$^{+1.38}_{-0.19}$ & 402 &      7.3$^{+\infty}_{-4.0}$   & 27.3   \\
Sk 82    & 0.06 & $139 \pm 7$  & 15.89$^{+0.36}_{-0.18}$ & 280 &      9.4$^{+3.4}_{-2.3}$      & 21.0   \\
AV 238   & 0.095& $151 \pm 13$ & 15.95$^{+2.53}_{-0.11}$ & 125 &      20.7$^{+\infty}_{-15.5}$ & 29.4   \\
AV 242   & 0.085& $154 \pm 10$ & 17.21$^{+1.00}_{-1.25}$ & 94  &      4.4$^{+4.5}_{-2.4}$      & 29.4   \\
AV 264   & 0.065& $170 \pm 22$ & 14.74$^{+0.40}_{-0.36}$ & 60  &      linear                   & 22.7   \\
AV 321   & 0.055& \nd          & $\leq 14.56$            & \nd &       \nd                     & 43.4   \\
AV 327   & 0.08 & $110 \pm 10$ & 14.79$^{+0.37}_{-0.26}$ & 104 &      linear                   & 45.0   \\
Sk 108   & 0.07 & \nd          & $\leq 14.56$            & \nd &      \nd                      & 47.1   \\
AV 372   & 0.115& $134 \pm 11$ & 16.55$^{+0.60}_{-0.24}$ & 108 &      14.9$^{+3.6}_{-3.4}$     & 19.5   \\
AV 378   & 0.085& $149 \pm 8$  & 14.92$^{+0.49}_{-0.31}$ & 42  &      linear                   & 35.9   \\
AV 388   & 0.105& $122 \pm 5$  & 19.40$^{+0.23}_{-0.20}$ & 63  &      3.3$^{+6.6}_{-2.8}$      & 17.3   \\
Sk 159   & 0.09 & $141 \pm 11$ & 18.94$^{+0.14}_{-0.29}$ & 84  &      4.9$^{+2.2}_{-1.8}$      & 22.5   \\
Sk 188   & 0.05 & $118 \pm 11$ & 15.70$^{+0.28}_{-0.12}$ & 91  &      6.0$^{+1.4}_{-1.3}$      & 11.3   \\
\enddata
\tablenotetext{a}{Color excess is total of Galactic plus SMC.
Corrections are applied to correlations in the text. } 
\tablenotetext{b}{LSR velocity assuming that the Galactic \h2\ lines lie at 
v$_{LSR}$ = 0 km s$^{-1}$.} 
\tablenotetext{c}{The tabulated N(H~I) are corrected for
the systematic effects discussed in \S~2.4. The values here are N(H~I) = (0.75 $\pm$ 0.25) N(21 cm).}
\end{deluxetable}
\normalsize


\begin{deluxetable}{lcccccccc}
\tablecolumns{9}
\tablenum{5}
\tablewidth{0pt}
\tablecaption{Rotational Level Populations\tablenotemark{1} - LMC Sight Lines}
\tablehead{
\colhead{Name}
&\colhead{N(0)}
&\colhead{N(1)}
&\colhead{N(2)}
&\colhead{N(3)}
&\colhead{N(4)}
&\colhead{N(5)}
&\colhead{N(6)}
&\colhead{$b_{\rm Dopp}$} \\ 
&\colhead{(cm$^{-2}$)}
&\colhead{(cm$^{-2}$)}
&\colhead{(cm$^{-2}$)} 
&\colhead{(cm$^{-2}$)} 
&\colhead{(cm$^{-2}$)} 
&\colhead{(cm$^{-2}$)} 
&\colhead{(cm$^{-2}$)} 
&\colhead{(km s$^{-1}$)} 
} 

\startdata
Sk -67 05  & 19.30$^{+0.05}_{-0.05}$ & 18.96$^{+0.06}_{-0.06}$ & 15.66$^{+0.32}_{-0.20}$ & 
                    15.29$^{+0.22}_{-0.16}$ & 14.62$^{+0.08}_{-0.07}$ & 
                    \nd & \nd& 6.2$^{+1.4}_{-1.1}$    \\  
Sk -67 14  & 14.18$^{+0.42}_{-0.34}$ & 14.86$^{+0.18}_{-0.14}$ & 
                    14.18$^{+0.28}_{-0.24}$ &  \nd  & \nd   & \nd   & 
                    \nd  & linear \\ 
HD 32109   & 18.40$^{+0.13}_{-0.18}$ & 18.33$^{+0.26}_{-0.37}$ & 
                    15.51$^{+1.79}_{-0.22}$ & 15.55$^{+1.86}_{-0.26}$ & 
                    14.93$^{+1.57}_{-0.13}$ & 14.58$^{+0.68}_{-0.07}$  & \nd  & 
                    5.8$^{+1.8}_{-3.6}$ \\  
Sk -65 21  & 17.81$^{+0.13}_{-0.41}$ & 17.94$^{+0.09}_{-0.27}$ & 
                    16.81$^{+0.40}_{-1.48}$ & 16.70$^{+0.43}_{-1.47}$ & 
                    14.23$^{+1.62}_{-0.15}$ & \nd   & \nd  &  2.3$^{+2.1}_{-1.7}$ \\ 
Sk -65 22  & 14.20$^{+0.17}_{-0.15}$ & 14.52$^{+0.17}_{-0.16}$ & 
                   14.23$^{+0.14}_{-0.12}$ & 14.28$^{+0.28}_{-0.21}$ & 
                    \nd  & \nd  & \nd  & linear  \\ 
HD 33133   & 17.48$^{+0.18}_{-1.06}$ & 16.00$^{+1.05}_{-0.82}$ & 
                    14.74$^{+0.76}_{-0.26}$ & 14.50$^{+0.50}_{-0.18}$ & 
                    \nd   & \nd   & \nd  & 3.5$^{+2.6}_{-1.4}$  \\ 
Sk -69 59  &$\leq$13.6 &$\leq$13.7 & \nd  & \nd  & \nd  & \nd  & \nd  & \nd \\
Sk -70 60  & $\leq$13.9 & $\leq$14.1 & \nd   &  \nd  & \nd   & \nd   & \nd  & \nd  \\ 
Sk -70 69  & $\leq$14.0 & $\leq$14.2 & \nd   &  \nd  & \nd   & \nd   & \nd  & \nd  \\ 
Sk -68 41  &$\leq$13.8 &$\leq$13.9 & \nd  & \nd  & \nd  & \nd  & \nd  & \nd \\ 
Sk -68 52  & 19.28$^{+0.05}_{-0.05}$ & 19.01$^{+0.05}_{-0.05}$ & 
              15.13$^{+1.72}_{-0.37}$ & 15.39$^{+1.70}_{-0.63}$ & 14.79$^{+1.90}_{-0.21}$ & 
              \nd  & \nd  & 4.9$^{+2.9}_{-3.6}$ \\  
Sk -67 69     & 14.82$^{+1.50}_{-0.13}$ & 15.13$^{+1.71}_{-0.15}$ & 
                    14.44$^{+0.34}_{-0.05}$ & 14.74$^{+0.85}_{-0.08}$ & 
                    14.04$^{+0.08}_{-0.03}$& \nd  & \nd  & 
                    6.9$^{+3.3}_{-4.0}$   \\ 
Sk -69 104 &$\leq$13.9  & $\leq$14.0 & \nd   &  \nd  & \nd   & \nd   & \nd  & \nd  \\ 
Sk -67 76  & 14.24$^{+0.40}_{-0.31}$ & 14.58$^{+0.31}_{-0.27}$ & 
              14.33$^{+0.44}_{-0.38}$ & 14.20$^{+0.50}_{-0.46}$ & 
              \nd  & \nd  & \nd  & linear \\ 
Sk -69 124 & $\leq$13.7 & $\leq$13.9 & \nd   &  \nd  & \nd   & \nd   & \nd  & \nd  \\ 
Sk -67 101    &$\leq$13.7 &$\leq$13.9 & \nd  & \nd  & \nd  & \nd  & \nd  & \nd \\
Sk -67 104    &$\leq$13.6 &$\leq$13.7 & \nd  & \nd  & \nd  & \nd  & \nd  & \nd \\
Sk -68 80     &$\leq$13.6 &$\leq$13.7 & \nd  & \nd  & \nd  & \nd  & \nd  & \nd \\ 
BI 170        &$\leq$13.7  & $\leq$13.9 & \nd   &  \nd  & \nd   & \nd   & \nd  & \nd  \\ 
Sk -67 111    &$\leq$14.3 &$\leq$14.5 & \nd  & \nd  & \nd  & \nd  & \nd  & \nd \\ 
BI 173        & 15.03$^{+0.40}_{-0.15}$ & 15.35$^{+0.53}_{-0.18}$ & 
                    14.58$^{+0.16}_{-0.09}$ & 14.70$^{+0.15}_{-0.08}$ & 
                    14.20$^{+0.06}_{-0.05}$ & \nd   & \nd  & 5.9$^{+2.2}_{-1.7}$  \\  
Sk -70 91     & 14.05$^{+0.38}_{-0.36}$ & 14.53$^{+0.22}_{-0.21}$ & 14.28$^{+1.00}_{-0.78}$ 
                  &  \nd  & \nd   & \nd   & \nd  & linear \\ 
Sk -66 100    &$\leq$13.9 &$\leq$14.0 & \nd  & \nd  & \nd  & \nd  & \nd  & \nd \\ 
HD 269582     &$\leq$13.8 &$\leq$13.9 & \nd  & \nd  & \nd  & \nd  & \nd  & \nd \\ 
HD 37026      & 15.01$^{+1.95}_{-0.31}$ & 15.15$^{+1.99}_{-0.26}$ & 14.27$^{+1.83}_{-0.09}$ & 
                    14.27$^{+1.44}_{-0.05}$ & \nd  & \nd  & \nd  & 
                    5.4$^{+3.0}_{-4.0}$   \\ 
Sk -71 45     & 18.06$^{+0.09}_{-0.09}$ & 18.26$^{+0.03}_{-0.03}$ & 
                    17.89$^{+0.17}_{-1.14}$ & 17.63$^{+0.15}_{-1.11}$ & 
                    16.61$^{+0.25}_{-1.05}$ & 15.85$^{+0.11}_{-0.09}$ & 
                    \nd  & 1.6$^{+2.8}_{-0.5}$ \\  
HD 269687     &$\leq$13.9  & $\leq$14.0 & \nd   &  \nd  & \nd   & \nd   & \nd  & \nd  \\ 
Sk -67 166    & 15.56$^{+1.15}_{-0.46}$ & 15.18$^{+1.25}_{-0.20}$ & 
                    14.31$^{+0.41}_{-0.08}$ & 14.20$^{+0.34}_{-0.07}$ & 
                    \nd   & \nd   & \nd  & 2.7$^{+0.7}_{-1.3}$ \\ 
Sk -67 169    &$\leq$13.6 &$\leq$13.7 & \nd  & \nd  & \nd  & \nd  & \nd  & \nd \\ 
Sk -67 167    &$\leq$13.8 &$\leq$14.0 & \nd  & \nd  & \nd  & \nd  & \nd  & \nd \\ 
Sk -67 191    &$\leq$13.9 &$\leq$14.0 & \nd  & \nd  & \nd  & \nd  & \nd  & \nd \\ 
BI 208        & 13.75$^{+0.47}_{-0.47}$ & 14.53$^{+0.22}_{-0.20}$ 
                      & \nd   &  \nd  & \nd   & \nd   & \nd  & linear \\ 
HD 37680      & 18.71$^{+0.03}_{-0.10}$ & 18.52$^{+0.07}_{-0.36}$ & 
                    16.87$^{+0.52}_{-1.16}$ & 17.11$^{+0.37}_{-1.53}$ &  
                    14.88$^{+0.41}_{-0.21}$ & 14.33$^{+0.15}_{-0.10}$ & 
                    \nd  & 3.7$^{+2.9}_{-1.4}$  \\
HD 269810     &$\leq$13.6 &$\leq$13.7 & \nd  & \nd  & \nd  & \nd  & \nd  & \nd \\ 
BI 229        &$\leq$13.7 &$\leq$13.9 & \nd  & \nd  & \nd  & \nd  & \nd  & \nd \\ 
Sk -66 169    & $\leq$13.7  & $\leq$13.8    & \nd  & \nd  & \nd  & \nd  & \nd  & \nd \\ 
Sk -66 172        & 18.15$^{+0.15}_{-0.27}$ & 17.32$^{+0.84}_{-1.05}$ & 
                    15.45$^{+1.94}_{-0.26}$ & 15.76$^{+1.76}_{-0.51}$ & 
		    14.50$^{+1.04}_{-0.08}$ & \nd   & \nd  & 6.0$^{+2.4}_{-4.1}$ \\  
Sk -68 135        & 19.47$^{+0.12}_{-0.12}$ & 19.61$^{+0.03}_{-0.03}$ & 
                    18.42$^{+0.09}_{-0.23}$ & 18.05$^{+0.11}_{-0.35}$ & 
                    17.17$^{+0.49}_{-1.19}$ & 16.91$^{+0.78}_{-1.15}$  & 14.73$^{+0.64}_{-0.15}$ & 
                    4.3$^{+2.4}_{-2.1}$ \\  
Sk -69 246    & 19.42$^{+0.02}_{-0.02}$ & 19.38$^{+0.03}_{-0.03}$ & 
                    17.75$^{+0.05}_{-1.19}$ & 17.83$^{+0.08}_{-1.44}$ & 
                    15.88$^{+0.31}_{-0.65}$ & 15.36$^{+0.18}_{-0.42}$ & 
                    14.88$^{+0.10}_{-0.27}$ & 4.8$^{+4.0}_{-0.5}$  \\ 
HD 269927         & 14.73$^{+0.09}_{-0.07}$ & 15.23$^{+0.30}_{-0.12}$ & 14.55$^{+0.15}_{-0.08}$ & 
                    14.87$^{+0.38}_{-0.13}$ & 14.37$^{+0.08}_{-0.05}$ & \nd  & \nd  & 
                    8.1$^{+3.9}_{-3.2}$  \\ 
BI 272            &$\leq$13.8 &$\leq$14.0 & \nd  & \nd  & \nd  & \nd  & \nd  & \nd \\ 
Sk -70 115        & 19.80$^{+0.07}_{-0.07}$ & 19.37$^{+0.07}_{-0.07}$ & 17.55$^{+0.34}_{-0.34}$   
                        & 17.12$^{+0.14}_{-0.81}$ & 
                                  15.40$^{+0.75}_{-0.43}$ & 14.80$^{+0.63}_{-0.24}$ & \nd
                                & 2.3$^{+1.4}_{-0.8}$ \\  
\enddata
\tablenotetext{1}{The 4 $\sigma$ upper limits are derived assuming 
statistical uncertainties and a 30 km s$^{-1}$ resolution element. } 
\end{deluxetable}


\begin{deluxetable}{lccccccc}
\renewcommand{\arraystretch}{1.4} 
\tablecolumns{9}
\tablenum{6}
\tablewidth{0pt}
\tablecaption{Rotational Level Populations\tablenotemark{1} - SMC Sight Lines}
\tablehead{
\colhead{Name}
&\colhead{N(0)} 
&\colhead{N(1)}
&\colhead{N(2)}
&\colhead{N(3)}
&\colhead{N(4)}
&\colhead{N(5)}
&\colhead{$b_{\rm Dopp}$} \\ 
&\colhead{(cm$^{-2}$)}
&\colhead{(cm$^{-2}$)}
&\colhead{(cm$^{-2}$)} 
&\colhead{(cm$^{-2}$)} 
&\colhead{(cm$^{-2}$)} 
&\colhead{(cm$^{-2}$)} 
&\colhead{(km s$^{-1}$)} 
} 

\startdata
AV 14             & 16.88$^{+1.24}_{-0.93}$ & 16.44$^{+1.59}_{-0.56}$ & 15.04$^{+0.28}_{-0.08}$ & 
                    15.21$^{+0.33}_{-0.09}$ & 14.65$^{+0.07}_{-0.05}$ & 
                    \nd   & 14.3$^{+5.6}_{-5.7}$ \\ 
AV 15             & 16.79$^{+1.14}_{-1.31}$ & 16.16$^{+1.21}_{-0.44}$ & 14.93$^{+0.24}_{-0.08}$ & 
                           15.16$^{+0.50}_{-0.13}$ & 14.50$^{+0.06}_{-0.03}$ & 
                           \nd  & 9.1$^{+3.0}_{-3.0}$  \\ 
AV 26             & 20.49$^{+0.04}_{-0.04}$ & 20.06$^{+0.07}_{-0.07}$ & 
                    17.99$^{+0.04}_{-1.58}$ & 17.91$^{+0.04}_{-1.56}$ & 
		    16.74$^{+0.45}_{-1.75}$ & \nd & 1.8$^{+5.9}_{-1.3}$ \\ 
AV 47             & 16.94$^{+0.51}_{-1.24}$ & 17.69$^{+0.27}_{-1.14}$  & 
                           15.79$^{+1.12}_{-0.54}$ & 15.30$^{+0.41}_{-0.29}$  & 
                           14.30$^{+0.06}_{-0.04}$ & \nd  & 6.1$^{+3.0}_{-2.0}$  \\ 
AV 69             & 18.36$^{+0.13}_{-0.13}$ & 18.49$^{+0.16}_{-0.16}$ & 15.89$^{+1.05}_{-0.40}$ & 
                           15.82$^{+1.07}_{-0.42}$ & 14.55$^{+0.07}_{-0.05}$ & 
                           \nd  & 7.2$^{+2.1}_{-2.2}$  \\ 
AV 75             & 18.42$^{+0.20}_{-0.58}$ & 17.80$^{+0.26}_{-0.88}$ & 15.56$^{+0.54}_{-0.28}$ &  
                           15.72$^{+0.63}_{-0.34}$ & 14.87$^{+0.11}_{-0.08}$ & 
                           14.75$^{+0.10}_{-0.09}$ & 9.6$^{+3.0}_{-2.1}$  \\ 
AV 83             & 14.25$^{+0.18}_{-0.18}$ & 14.61$^{+0.06}_{-0.06}$ & 
                    14.49$^{+0.07}_{-0.07}$ & 14.74$^{+0.14}_{-0.12}$ & 
                    \nd   & \nd   & linear \\ 
AV 95             & 19.09$^{+0.09}_{-0.09}$ & 19.10$^{+0.06}_{-0.06}$ & 15.30$^{+0.25}_{-0.18}$ & 
                    15.20$^{+0.19}_{-0.14}$ & 14.47$^{+0.02}_{-0.02}$ & \nd   
                    & 10.3$^{+3.9}_{-2.0}$  \\ 
AV 207            & 19.09$^{+0.10}_{-0.10}$ & 19.09$^{+0.06}_{-0.06}$ & 
                      17.47$^{+0.17}_{-2.09}$ & 17.16$^{+0.26}_{-2.09}$ & 
                      14.91$^{+1.26}_{-0.46}$ & \nd   & 
		      2.7$^{+6.1}_{-1.2}$  \\ 
NGC 346-3         & 15.33$^{+1.66}_{-0.14}$ & 15.08$^{+0.32}_{-0.08}$ & 
                    14.66$^{+0.14}_{-0.05}$ & 14.61$^{+0.15}_{-0.05}$ & 
                    \nd   & \nd   & 17.8$^{+\infty}_{-9.7}$ \\ 
NGC 346-4         & 14.87$^{+0.49}_{-0.08}$ & 15.00$^{+0.74}_{-0.07}$ & 
                    14.42$^{+0.24}_{-0.03}$ & 14.53$^{+0.13}_{-0.03}$ & 
		    \nd   & \nd   & 19.0$^{+\infty}_{-11.3}$ \\ 
NGC 346-6         & 15.19$^{+0.40}_{-0.12}$ & 15.44$^{+0.52}_{-0.12}$ & 
                    14.77$^{+0.10}_{-0.05}$ & 14.82$^{+0.14}_{-0.06}$ & 
		    \nd   & \nd   & 17.3$^{+5.9}_{-7.4}$  \\ 
HD 5980           & 15.14$^{+0.38}_{-0.19}$ & 15.36$^{+0.24}_{-0.14}$ & 
                    14.70$^{+0.04}_{-0.04}$ & 14.59$^{+0.04}_{-0.03}$ & \nd  & \nd  & 
                    9.2$^{+3.0}_{-2.0}$ \\ 
AV 232            & 14.26$^{+0.26}_{-0.10}$ & 15.03$^{+1.60}_{-0.29}$ & 
                    14.38$^{+0.13}_{-0.07}$ & 14.51$^{+0.27}_{-0.09}$ & \nd  & \nd  & 
                    7.3$^{+\infty}_{-4.0}$  \\ 
Sk 82             & 14.95$^{+0.16}_{-0.14}$ & 15.64$^{+0.48}_{-0.23}$ & 15.01$^{+0.15}_{-0.13}$ & 
                    15.09$^{+0.19}_{-0.12}$ & 14.34$^{+0.04}_{-0.04}$ & 
                    \nd & 9.4$^{+3.4}_{-2.3}$ \\ 
AV 238            & 15.37$^{+2.56}_{-0.09}$ & 15.73$^{+2.60}_{-0.13}$ & 
                    14.81$^{+0.69}_{-0.04}$ & 14.72$^{+0.49}_{-0.03}$ & 
		    \nd   & \nd   & 20.7$^{+\infty}_{-15.5}$ \\ 
AV 242            & 16.80$^{+0.98}_{-1.36}$ & 16.97$^{+0.94}_{-1.27}$ & 
                    15.48$^{+1.58}_{-0.64}$ & 15.27$^{+1.63}_{-0.50}$ & \nd   & \nd & 
                         4.4$^{+4.5}_{-2.4}$ \\ 
AV 264            & 14.55$^{+0.42}_{-0.40}$ & 14.28$^{+0.36}_{-0.31}$ & 
                    \nd   & \nd   & \nd   & \nd   & linear \\ 
AV 321            &$\leq$14.0 &$\leq$14.2 & \nd   & \nd   & \nd   & \nd   & \nd   \\ 
AV 327     & 14.21$^{+0.18}_{-0.17}$ & 14.45$^{+0.51}_{-0.34}$ & 
                    14.23$^{+0.22}_{-0.23}$ & \nd   & \nd   & \nd   & 
		    linear \\ 
Sk 108            &$\leq$14.0 &$\leq$14.2 & \nd  & \nd  & \nd  & \nd  & \nd  \\ 
AV 372            & 16.01$^{+0.56}_{-0.18}$ & 16.28$^{+0.69}_{-0.29}$ & 
                    15.45$^{+0.22}_{-0.13}$ & 15.40$^{+0.29}_{-0.12}$ & 
                    14.56$^{+0.05}_{-0.04}$ & 14.60$^{+0.05}_{-0.04}$ 
                    & 14.9$^{+3.6}_{-3.4}$ \\ 
AV 378            & 14.22$^{+0.58}_{-0.45}$ & 14.48$^{+0.22}_{-0.17}$ & 
                    14.15$^{+0.84}_{-0.70}$ & 14.34$^{+0.30}_{-0.27}$  & 
		    \nd   & \nd   & linear  \\ 
AV 388            & 19.19$^{+0.11}_{-0.11}$ & 18.96$^{+0.38}_{-0.38}$ & 
                    17.63$^{+0.16}_{-2.25}$ & 17.24$^{+0.49}_{-1.87}$ &  
		    15.79$^{+1.31}_{-1.11}$ & \nd   & 3.3$^{+6.6}_{-2.8}$ \\ 
Sk 159            & 18.60$^{+0.13}_{-0.29}$ & 18.67$^{+0.13}_{-0.29}$ & 
                    16.60$^{+0.80}_{-0.88}$ & 16.13$^{+1.12}_{-0.80}$ & 
		    \nd   & \nd   & 4.9$^{+2.2}_{-1.8}$  \\ 
Sk 188            & 15.23$^{+0.34}_{-0.14}$ & 15.37$^{+0.30}_{-0.14}$ & 
                    14.76$^{+0.10}_{-0.06}$ & 14.57$^{+0.08}_{-0.04}$ & 
                    \nd   & \nd   & 6.0$^{+1.4}_{-1.3}$  \\
\enddata
\tablenotetext{1}{The 4 $\sigma$ upper limits are dervied assuming 
statistical uncertainties and a 30 km s$^{-1}$ resolution element.} 
\end{deluxetable}
\normalsize


\end{document}